\documentclass[fleqn,usenatbib]{mnras}

\usepackage{newtxtext,newtxmath}

\usepackage[T1]{fontenc}
\usepackage{comment}

\DeclareRobustCommand{\VAN}[3]{#2}
\let\VANthebibliography\thebibliography
\def\thebibliography{\DeclareRobustCommand{\VAN}[3]{##3}\VANthebibliography}

\usepackage{graphicx}	
\usepackage{amsmath}	

\title[Short title, max. 45 characters]{Dust and Gas Transport in Substructured Nonideal MHD Wind-Launching Disks with Embedded Planets}

\author[C.-Y. Hsu et al.]{
Chun-Yen Hsu,$^{1,2}$\thanks{E-mail: kdj8qp@virginia.edu}
Zhi-Yun Li,$^{1,2}$
Xiao Hu,$^{1,3}$
Yisheng Tu,$^{1,2}$
Min-Kai Lin,$^{4,5,6}$
\\
$^{1}$Department of Astronomy, University of Virginia, Charlottesville, VA 22904, USA\\
$^{2}$ Virginia Institute of Theoretical Astronomy, University of Virginia, Charlottesville, VA 22904, USA\\
$^{3}$Department of Astronomy, University of Florida, Gainesville, FL 32608, USA\\
$^{4}$Institute of Astronomy and Astrophysics, Academia Sinica, Taipei 10617, Taiwan\\
$^{5}$Physics Division, National Center for Theoretical Sciences, Taipei 10617, Taiwan \\
$^{6}$Tsung-Dao Lee Institute, Shanghai Jiao Tong University, Shanghai, P.R.C.
}

\date{Accepted 2026 July 20. Received 2026 July 6; in original form 2026 April 20}

\pubyear{\the\year{}}

\begin{document}
\label{firstpage}
\pagerange{\pageref{firstpage}--\pageref{lastpage}}
\maketitle

\begin{abstract}

Radial dust transport in protoplanetary disks is a key process shaping planet formation and disk chemistry. We investigate how this transport, along with gas transport, is regulated in wind-launching disks with embedded planets using three-dimensional nonideal MHD simulations. We find that disk substructures do not act as absolute barriers to transport. Low-mass planets leave the disk structure dominated by the magnetic wind, while a Jupiter-mass planet opens a deep gap and drives spiral shocks. However, even in this regime, wind-driven accretion persists; the planet reshapes rather than replaces the magnetically driven flow, leaving the gap intrinsically time-dependent and partially permeable. Early-phase suppression of inward transport is followed by the development of localized, azimuthally intermittent inflow channels that enable continued cross-gap transport. This transport is strongly size-dependent: small grains remain coupled to the gas and readily penetrate the gap, whereas larger grains are efficiently trapped outside the planet. Consequently, a giant planet acts as an efficient but incomplete filter rather than a perfect barrier. These results support a "leaky gap" scenario, where radial transport is regulated rather than halted by substructures. Volatile-rich material can be delivered to the inner disk both before gap opening and via continued leakage, providing a natural explanation for the diverse inner disk compositions inferred from JWST. Similarly, pebble isolation during core growth should be viewed as a gradual filtering process rather than a binary transition. More generally, disk substructures are dynamically evolving features whose transport efficiency depends on their physical origin (magnetic versus planet-driven).

\end{abstract}




\section{Introduction}

The formation and evolution of planetary systems are intimately linked to the dynamics of protoplanetary disks, the birthplace of planets. These accretion disks, composed primarily of gas and dust, undergo complex physical processes that determine the initial conditions for planet formation and the resulting planetary architectures. A key aspect of disk evolution is the redistribution of mass and angular momentum, which is strongly influenced by magnetic fields. In particular, non-ideal magnetohydrodynamic (MHD) wind-launching disks have emerged as a leading paradigm, in which angular momentum removal by disk winds drives accretion onto the central star and shapes the disk structure \citep[e.g.,][]{Wang19,Cui21,Aoyama23}.

Within these evolving disks, the radial transport of both gas and solids plays a central role in regulating disk chemistry and planet formation. Dust particles experience aerodynamic coupling with the gas, which regulates their radial drift, vertical settling, and concentration within substructures. The inward transport of solids, especially millimeter- to centimeter-sized grains, provides a key pathway for redistributing mass and volatile material from the outer disk to the inner regions. This redistribution underlies a wide range of processes, including the delivery of volatile-rich material to the terrestrial planet-forming zone and the buildup of planetary cores.

The presence of embedded planets introduces a critical modification to this transport. Growing planets can carve gaps in the disk, altering both the gas flow and the distribution of solids. In hydrodynamic models, such substructures are known to partially impede radial transport, leading to size-dependent dust filtration and reduced but non-zero mass flux across gaps \citep[e.g.,][]{Lubow05,Bitsch18,Huang25c}. However, most previous studies have been conducted in viscous or turbulent hydrodynamic disks, where angular momentum transport is typically modeled through local turbulent stresses parameterized as an effective viscosity, and it remains unclear how radial transport operates in magnetically driven, wind-launching disks, where angular momentum transport and accretion are controlled by large-scale magnetic stresses rather than local viscosity.

It is important to note that disk substructures are not uniquely produced by planets. In non-ideal MHD disks, magnetic processes can also generate rings and gaps through spatial variations in magnetic stress, magnetic flux redistribution, and wind-driven mass loss \citep[e.g.,][]{Bethune17,Suriano18,Riols20,Cui21,Hsu24}. This raises a key question: do substructures of different physical origins regulate the transport of gas and solids in similar ways, or does their ability to filter material depend fundamentally on whether they are magnetically driven or planet-induced?

Recent three-dimensional non-ideal MHD simulations have further shown that embedded planets can significantly modify the magnetic structure of disks, concentrating magnetic flux in the planet-opened gap and altering the efficiency of angular momentum removal there \citep[e.g.,][]{Aoyama23,Wafflard23,Hu25}. These effects can deepen planetary gaps and reshape the global accretion flow. Whether such gaps act as effective barriers to radial transport or remain partially permeable remains an open question.

The above considerations suggest that disk substructures may not act as strict barriers to radial transport but rather as filters whose efficiency depends on their physical origin. Magnetically generated substructures are expected to be highly permeable, reflecting their prominent magnetically driven accretion streams and meridional circulation patterns \citep{Hu22}, while gaps opened by massive planets may impose stronger constraints on transport. Quantifying this ``leaky gap'' behavior, and its dependence on the origin of substructures, is essential for understanding how gas and solids are redistributed in planet-forming disks.

This question has important implications for both disk chemistry and planet formation. The radial redistribution of solids and volatiles influences the chemical composition of the inner disk, including key diagnostics such as the carbon-to-oxygen (C/O) ratio, which is now being probed by facilities such as ALMA and JWST \citep[e.g.,][]{Andrews18,Oberg21,Pascucci25}. It also affects the extent to which different regions of the disk remain chemically connected or become isolated, with potential links to the dichotomy between non-carbonaceous and carbonaceous meteorites in the early Solar System \citep[e.g.,][]{Spitzer25}. In addition, the efficiency of radial transport regulates the supply of solids available for planetary growth, providing the broader context within which processes such as pebble accretion (the preferential accretion of drifting pebble-sized solids by growing planetary cores \citep[e.g.,][]{Lambrechts12}) operate.

In this paper, we investigate how the radial transport of gas and dust is regulated in non-ideal MHD wind-launching disks with embedded planets using three-dimensional non-ideal MHD simulations. Our focus is on the global transport of material across both planet-induced and MHD-generated substructures and the extent to which these structures act as barriers or filters. We explore how wind-driven accretion, magnetic stresses, and planet--disk interactions jointly determine the permeability of gaps and the resulting redistribution of solids.
Our results show that disk substructures do not act as absolute barriers to transport but instead regulate it in a time-dependent and size-dependent manner, with an efficiency that strongly depends on their physical origin (magnetic vs planet-driven). These findings provide a new framework for interpreting disk substructures as dynamically evolving filters rather than static barriers, with implications for volatile delivery, disk chemistry, and planet growth.

The paper is organized as follows. Section~\S\ref{sec:simulation} introduces the governing equations, simulation setup, and numerical methods. Section~\S\ref{sec:gas_dynamics_transport} presents the gas dynamics and transport in disks with embedded planets. Section~\S\ref{sec:dust_dynamics_transport_feedback} includes dust dynamics, focusing on radial transport and size-dependent behavior. Section~\S\ref{sec:discussion} discusses the implications of our results for disk chemistry and planet formation. Our main conclusions are summarized in Section~\S\ref{sec:conclusion}.
\section{Simulation} 
\label{sec:simulation}

We use Athena++ \citep{stone20} with a pressureless dust fluid module  \citep{Huang22,Huang25}  to solve the non-ideal MHD gas and dust fluid equations:
\begin{equation}
  \frac{\partial \rho_{\rm g}}{\partial t} + \nabla \cdot (\rho_{\rm g} {\rm \bf V}_{\rm g}) = 0, \label{eq:gas_continuity}
\end{equation}
\begin{equation}
  \frac{\partial (\rho_{\rm g} {\rm \bf V}_{\rm g})}{\partial t} + \nabla \cdot (\rho_{\rm g} {\rm \bf V}_{\rm g} {\rm \bf V}_{\rm g} + {\rm P^*} {\rm {\bf I}}  - \frac{{\rm \bf B} {\rm \bf B} }{4 \pi}) = - \rho_{\rm g} \nabla \Phi + \rho_{\rm d} \frac{{\rm \bf V}_{\rm d} - {\rm \bf V}_{\rm g}}{\tau_{\rm s}}, \label{eq:gas_momentum}
\end{equation}
\begin{align}
  \frac{\partial E_{\rm g}}{\partial t} + \nabla \cdot  \left[(E_{\rm g} + {\rm P^*}){\rm \bf V}_{\rm g} - \frac{{\rm \bf B} ({\rm \bf B} \cdot {\rm \bf V}_{\rm g})}{4 \pi} + \frac{1}{c} \left( \eta_O {\rm \bf J} + \eta_{\rm AD} {\rm \bf J}_{\perp} \right) \times {\rm \bf B}  \right] \notag \\
  = - \rho_{\rm g} ({\rm \bf V}_{\rm g} \cdot \nabla \Phi) - \Lambda_{\rm c} + \rho_{\rm d} \frac{{\rm \bf V}_{\rm d} - {\rm \bf V}_{\rm g}}{\tau_{\rm s}} \cdot {\rm \bf V}_{\rm g} + \omega_{\rm d} \rho_{\rm d} \frac{|{\rm \bf V}_{\rm d} - {\rm \bf V}_{\rm g}|^2}{\tau_{\rm s}}, \label{eq:gas_energy}
\end{align}
\begin{equation}
\frac{\partial \rho_{\rm d}}{\partial t} + \nabla \cdot (\rho_{\rm d} {\rm \bf V}_{\rm d} ) = 0, \label{eq:dust_continuity}
\end{equation}
\begin{equation}
\frac{\partial \rho_{\rm d} ({\rm \bf V}_{\rm d})}{\partial t} + \nabla \cdot (\rho_{\rm d} {\rm \bf V}_{\rm d} {\rm \bf V}_{\rm d}) = - \rho_{\rm d} \nabla \Phi + \rho_{\rm d} \frac{{\rm \bf V}_{\rm g} - {\rm \bf V}_{\rm d}}{\tau_{\rm s}}, \label{eq:dust_momentum}
\end{equation}
and the induction equation
\begin{equation}
  \frac{\partial {\rm \bf B}}{\partial t} = \nabla \times ({\rm \bf V}_{\rm g} \times {\rm \bf B}) - \frac{4 \pi}{c} \nabla \times \left[ \eta_O {\rm \bf J} + \eta_{\rm AD} {\rm \bf J}_{\perp} \right] , \label{eq:induction}
\end{equation}
where $\rho_{\rm g}$ and ${\rm \bf V}_{\rm g}$ are gas mass density and velocity, $\rho_{\rm d}$ and ${\rm \bf V}_{\rm d}$ are dust mass density and velocity, ${\rm \bf B}$ is the magnetic field, ${\rm P^*} = P_{\rm g} + B^2/(8\pi)$ is the total (thermal [$P_{\rm g}$] and magnetic)  pressure, ${\rm \bf I}$ is the identity tensor, $\Phi = -GM/r$ is the gravitational potential of the central star, $E_{\rm g} = \rho_{\rm g} V_{\rm g}^2/2 + {P_{\rm g}}/(\gamma - 1) + B^2/(8 \pi)$ is the energy density, $\gamma$ is the adiabatic index, ${\rm \bf J}$ is the current density, ${\rm \bf J}_{\perp} = {\rm \bf B} \times ({\rm \bf J} \times {\rm \bf B}) / (B^2)$ is the component of ${\rm \bf J}$ perpendicular to the magnetic field, $\eta_O$ and $\eta_{\rm AD}$ are the Ohmic and ambipolar diffusivities, and $\Lambda_{\rm c}$ is the cooling term.
The aerodynamic drag between gas and dust is included in the last terms of Eq. \ref{eq:gas_momentum} and \ref{eq:dust_momentum}. The dust fluid module in Athena++ assumes linear drag law, so the stopping time $\tau_{\rm s}$ is independent of velocity.
The last two source terms on the right hand side of the gas energy equation (Eq.~\ref{eq:gas_energy}) are drag and frictional heating terms, respectively. The parameter $\omega_{\rm d}$ is used to control the level of frictional heating, and we set $\omega_{\rm d} = 1$ to assume all the dissipation is deposited in the gas. Although we solve the full energy equation (Eq. 3), we employ a rapid cooling scheme that keeps the disk close to locally isothermal during the evolution. Our simulations are conducted in a corotating frame centred on the embedded planet (see Appendix A of \citealt{Aoyama23} for details). We present the azimuthal velocity in the corotating frame in this work unless otherwise noted.

\subsection{Simulation domain} \label{subsection:simulation_domain}

We perform fully 3D simulations using spherical coordinates ($r, \theta, \phi$). The simulation domain spans 1 to 316~au radially, 0.05 to $\pi$-0.05 in the polar direction, and 0 to $2\pi$ azimuthally. The radial grid follows a logarithmic spacing with 80 base cells and an adjacent cell size ratio of 1.0746. The polar grid is uniform with 96 base cells and uniform with 32 base cells in the azimuthal direction. The base cells in the simulation domain are the same as those used in \cite{Hsu25}.
We apply five levels of static mesh refinement (SMR), with each level refining the grid by a factor of 2. The finest level covers the grids near the embedded planet at 10~au, extends radially from 9.3 to 10.7 au, spans approximately one scale height (equivalent to ~0.05 radians) above and below the mid-plane, and azimuthally, this level covers $\phi$ from -0.05 to +0.05 radians around $\phi=\pi$, where the planet is located. The fourth refinement level covers the grids near the embedded planet at 10~au, extending radially from 9.0 to 11.0 au, spans approximately 1.4 scale heights (equivalent to ~0.07 radians) above and below the mid-plane, and azimuthally, this level covers $\phi$ from -0.1 to +0.1 radians around $\phi=\pi$. We apply a third refinement level that extends radially from 2.37 to 100 au, spans approximately 2.5 scale heights (equivalent to ~0.13 radians) above and below the mid-plane, and resolves the disk scale height with approximately 12.6 grid cells. Azimuthally, this level covers the entire $\phi$ domain. The second and first refinement levels cover, respectively, 5.57-176 au and 3.14–316 au radially, and $-$0.25 to $+$ 0.25 radians and $-$0.5 to $+$ 0.5 radians of the mid-plane in the polar direction.

\subsection{Boundary conditions} 
\label{subsection:boundary conditions}

We implemented modified outflow boundary conditions at both the inner and outer radial boundaries. At the inner boundary, instead of copying the gas density and pressure from the innermost active zone to the ghost zones, we extended these quantities using the same power-law profile applied during gas initialization. For gas velocity, the azimuthal component follows a Keplerian profile, while the radial and polar components are copied from the innermost active zone, ensuring that no external mass enters the simulation domain. Reflective boundary conditions are applied in the polar ($\theta$) direction, and periodic boundary conditions are used in the azimuthal direction. We adopt reflective boundary conditions in the polar direction primarily to prevent magnetic flux from artificially escaping through the polar boundaries and to maintain numerical stability. Since the primary focus of this work is the disk region near the midplane and far from the polar boundary, this choice is not expected to qualitatively affect the transport behavior studied here.

\subsection{Initial conditions} \label{initial_conditions_subsection}

We perform simulations initialized in hydrostatic equilibrium, with a magnetic field derived from the vector potential to ensure a divergence-free $B$ field (i.e. $\nabla \cdot {\rm \bf B} = 0$) and initially axisymmetric setup in the azimuthal direction.

 The initial conditions of the gas and the magnetic field are the same as in \cite{Hsu24}, and the initial temperature and gas density profiles follow those used in \cite{Hu22}. Specifically, we divide the simulation domain into a cold and dense disk and a hot low-density corona, maintaining a constant aspect ratio of $h/r=0.05$,  where $h$ is the height of the disk scale. The cold and dense disk is confined within two scale heights above and below the midplane, defined as $\pi/2 - \theta_0 <\theta < \pi/2 + \theta_0$, where $\theta_0=\arctan{(2h/r)}$. The gas density and midplane temperature both follow power-law distributions with indices $p=-1.5$ and $q=-1$, respectively:
\begin{eqnarray}
\rho(r,\pi/2) = \rho_0(r/r_0)^{p},  \\ \nonumber 
T(r,\pi/2) = T_0(r/r_0)^{q} 
\label{eq:profile}
\end{eqnarray}
where $r_0=1$~au is the radius of the inner boundary of the computational domain, and $\rho_0=2.667 \times 10^{-10} {\rm g/cm^{-3}}$ and $T_0=570$~K are the density and temperature at $r_0$. To ensure a smooth transition between the cold disk and hot corona, we adopt the following vertical profile for the temperature:  
\begin{equation}
T(r,\theta)=
\begin{cases}
T(r,\pi/2) & \text{if }  |\theta-\pi/2| < \theta_0 \\
T(r,\pi/2)\ {\rm exp}[(|\theta-\pi/2|\\
\ -\theta_0)/\theta_0 \times\ln(160)]
  & \text{if } \theta_0 \leq |\theta-\pi/2| \leq 2\theta_0  \\
160\ T(r,\pi/2); & \text{if } |\theta-\pi/2| > 2\theta_0\\
\end{cases}\label{eq:T}
\end{equation}
We use a quick $\beta_{\rm cool}$ cooling scheme with a cooling timescale of only $10^{-10}$ of the local orbital period, so the temperature profile is effectively fixed over time.
The vertical density profile is generated based on hydrostatic equilibrium, i.e.,$v_r=v_\theta=0$. 

The initial magnetic field is computed from the magnetic vector potential used in \cite{Zanni07}:
\begin{align}
 & B_r (r, \theta) = \frac{1}{r^2 {\rm sin}\theta} \frac{\partial A_\phi}{\partial\theta}, \label{eq:Br} \\
 & B_\theta (r, \theta) = -\frac{1}{r\ {\rm sin}\theta} \frac{\partial A_\phi}{\partial r}, \label{eq:Btheta} \\
 & B_\phi = 0, \label{eq:Bphi}
\end{align}
with
\begin{align}
 & A_\phi (r, \theta) = \frac{4}{3}r_0^2 B_{\rm p,0} \left( \frac{r {\rm sin}\theta}{r_0}\right)^{\frac{3}{4}} \frac{1}{(1 + 2{\rm cot^2 \theta})^{5/8}}
\end{align}
where $B_{\rm p,0}$ sets the scale for the poloidal field strength. The magnetic field setup is the same as that of \cite{Bai17}, \cite{Suriano18}, and \cite{Hu22}. In all our simulations, $B_{\rm p,0}$ is set by plasma-$\beta = 10^3$ in the mid-plane. 

We set the initial total dust mass density $\rho_{\rm d}$ to $1\%$ of the initial gas mass density $\rho_{\rm g}$ and the initial dust velocity the same as that of the gas. In the relatively dense regions within four gas scale heights of the midplane, we computed the dust stopping time based on the dust size $a$:
\begin{align}
 &\tau_{\rm s} = \frac{\rho_{\rm m}}{\rho_{\rm g}} \frac{a}{v_{\rm th}}
\end{align}
where $\rho_{\rm m}$ is the grain material density, taken to be 3~${\rm g/cm^{3}}$ here and in the chemistry network (\S \ref{subsec: chemistry and non-ideal}). $v_{\rm th}$ is the thermal speed of the gas. In the low-density regions above and below four gas scale heights from the midplane, we fix the Stokes number ${\rm St}=\tau_{\rm s}\Omega_{\rm K}$ at $0.01$ to make the simulation run more stably. This assumption does not significantly affect our conclusions since they will be based primarily on grains that settle well below four gas scale heights.  

\subsection{Planet} \label{subsec: planet}
The planet is placed on a fixed circular orbit at $r=$10~au in the disc mid-plane \footnote{We note that the substructures can be affected by planetary migration \citep[e.g.,][]{Kanagawa20} and orbital eccentricity \citep[e.g.,][]{Salcedo23,Pichierri24}.}.
The total gravitational potential in the rotation frame is given as
\begin{align}
    & \Phi = - GM_{\star} \left[ \frac{1}{r} + \frac{M_{\rm p}/M_{\star}}{\sqrt{|{\bf r} - { R}_{\rm p}|^2+r_{\rm s}^2}}\right]
\end{align}
where $M_{\star}$ and $M_{\rm p}$ are the masses of the central star and planet, respectively, $R_{\rm p}$ is the orbital radius of the planet, and $r_{\rm s}$ is the softening radius to avoid the gravitational potential approaching infinite near the planet. We set $M_{\star}=1M_\odot$ and $r_{\rm s}=0.04$~au to be 2 times the smallest cell size in the Hill sphere of the planet. We ignore the indirect potential term due to the offset between the central star and the center of gravity of the star–planet system, as it has a negligible impact on the perturbation structure \citep{Miranda19,Hu25}. 

The Hill radius of a planet is defined as \citep{Armitage24}
\begin{align}
 & r_{\rm H} = \left(\frac{M_{\rm p}}{3M_{\star}} \right)^{\frac{1}{3}}R_{\rm p}
\end{align}
where $r_{\rm H} \sim $ 0.693~au in our Models with 1 Jupiter mass of the embedded planet, and it is larger than $h_{\rm p} =$ 0.5~au \footnote{We note that the circumplanetary disk and gap substructures formation require a sufficiently large planet mass $M_{\rm p}$ relative to the thermal mass $M_{\rm th}$ (roughly, $r_{\rm H}>h_{\rm p}$) \citep[e.g.,][]{Sagynbayeva24,Armitage24}.}. We have ($r$, $\theta$, $\phi$) = (62, 140, 22) cells for the grids in the planetary Hill sphere in the Models with 1 Jupiter mass of the embedded planet.

\subsection{Chemical network and the non-ideal MHD coefficients} 
\label{subsec: chemistry and non-ideal}

The chemical network and the non-ideal MHD coefficients used in this work are the same as in \cite{Hsu24} (see their Sec. 3 and Appendix A for details). Briefly, we used a simplified chemical network following \cite{Umebayashi90}, \cite{Nishi91}, and \cite{grassi19}, which include the element H, He, C, O, and Mg, an MRN-type \citep{Mathis77}  power-law grain size distribution (with $a_{\rm min} = 0.5$ $\mu$m and $a_{\rm max} = 25$ $\mu$m, and a power-law index of $-3.5$), and gas-phase, gas-grain, and grain-grain reactions. Once the charge abundances are computed from the chemical network, they are stored in a lookup table and referenced during simulations to calculate the Ohmic and ambipolar diffusivities for each computational cell using the standard formulae (see, e.g., Eq.~[18]-[24] of \citealt{Hsu24}).  

To mimic the enhanced magnetic coupling (and thus reduced diffusivity) expected from increased ionization by stellar UV and X-ray irradiation in the lower-density regions near the disk surface and in the disk wind, we follow \cite{Suriano18} and multiply the computed Ohmic and ambipolar diffusivities by the following $\theta$ dependence: 

\begin{align}
 & f(\theta) = 
\begin{cases} 
{\rm exp} \left(-\frac{{\rm cos}^2(\theta + \theta_0)}{2(h/r)^2} \right) & \mbox{if} \quad  \theta < \frac{\pi}{2} -  \theta_0 \\
1 & \mbox{if} \quad \frac{\pi}{2} - \theta_0 \leq \theta   \leq \frac{\pi}{2} + \theta_0 \\
{\rm exp} \left(-\frac{{\rm cos}^2(\theta - \theta_0)}{2(h/r)^2} \right) & \mbox{if} \quad \theta > \frac{\pi}{2} + \theta_0 
\end{cases} .
\label{eq:theta_depend_non_ideal} 
\end{align}
In our setup, ambipolar diffusion dominates over Ohmic dissipation throughout most of the simulation domain, especially in the outer disk and in the low-density disk atmosphere. Near the dense inner-disk midplane, the two diffusivities become more comparable, although ambipolar diffusion generally remains similar to or larger than Ohmic dissipation. Angular momentum transport in the simulations is mediated primarily by large-scale magnetic stresses associated with the magnetized disk wind, especially outside the orbit of the planet.

\section{Gas dynamics and transport without dust} 
\label{sec:gas_dynamics_transport}

\begin{table*}
\caption {
Models
\label{table_cases}
} 
\begin{tabular}{llllll}
\hline
  Name  & planet's mass& Dust fluid & Dust fluid number  & Initial $\epsilon$ & size distribution  \\
        \hline
    M001J   & 0.01 $M_{\rm J}$& $\times$ & $\times$ & $\times$ & $\times$ \\    
    M01J   & 0.1 $M_{\rm J}$& $\times$ & $\times$ & $\times$ & $\times$ \\
    M1J   & 1 $M_{\rm J}$& $\times$ & $\times$ & $\times$ & $\times$ \\
    M001JD10mm01mm5bins & 0.01 $M_{\rm J}$& $\bigcirc$ & 5 & 0.01 &MRN \\
    M01JD10mm01mm5bins  & 0.1 $M_{\rm J}$& $\bigcirc$ & 5 & 0.01 &MRN \\
    M1JD10mm01mm5bins  & 1 $M_{\rm J}$& $\bigcirc$ & 5 & 0.01 &MRN \\
    \hline
\end{tabular}
\end{table*}

Table~\ref{table_cases} summarizes the models used in this study. Models M1J, M01J, and M001J represent 3D gas-only simulations with an embedded planet of mass 1, 0.1, and 0.01 Jupiter masses, respectively, located at 10~au. These models isolate the interaction between an embedded planet and a non-ideal MHD wind-launching disk. The corresponding simulations that include multiple dust bins are discussed in Section~\ref{sec:dust_dynamics_transport_feedback}.

In this section, we first characterize the global morphology of the gas disk and its dependence on the planet's mass. We then examine how the three-dimensional flow structure differs between the models and, finally, quantify the radial gas transport, particularly near the planet’s orbit. The goal is to establish the gas-dynamical framework within which dust dynamics and transport will later be interpreted.

\subsection{Global morphology and gap formation}
\label{subsec:gas_morphology}

\begin{figure*}
   \centering
\includegraphics[width=\linewidth]{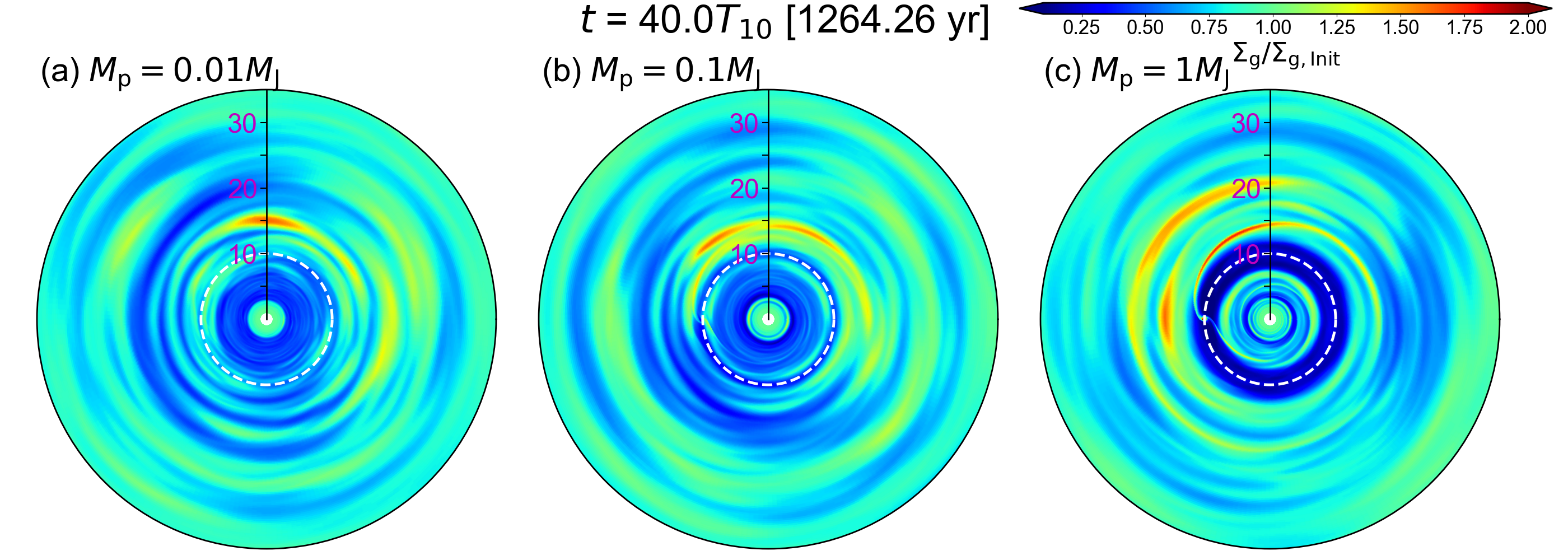}
    \caption{Gas surface density distributions at $t = 40\,T_{10}$ for disks hosting embedded planets of mass (a) $0.01\,M_{\rm J}$, (b) $0.1\,M_{\rm J}$, and (c) $1\,M_{\rm J}$ at $(r, \theta, \phi) = (10, \pi/2, \pi)$. The white dashed circles correspond to $r$ = 10~au. The lowest-mass planet produces only weak perturbations, leaving the wind-driven ring–gap structure largely intact. The intermediate-mass case exhibits a mixed morphology shaped by both magnetic stresses and planet-induced spirals. The Jupiter-mass planet opens a deep, wide gap and excites prominent large-scale spiral density waves. Dense arcs visible in some panels are associated with Rossby wave instability. The displayed snapshots correspond to representative late-time morphologies of each model. An animated version is available at \url{https://figshare.com/s/4a74041834cc33597ab0}.}
   \label{fig: pure_gas_face_on_view}
\end{figure*}

We begin by examining the surface density structure of the gas-only simulations. Figure~\ref{fig: pure_gas_face_on_view} shows the disk surface density at $t = 40\,T_{10}$, where $T_{10}$ is the orbital period at 10~au (corresponding to approximately 1200 years). The displayed snapshots correspond to representative late-time states of the simulations after major gas substructures have developed. The accompanying animations are included primarily to illustrate the time evolution and temporal variability of these structures. 

In the lowest-mass case ($M_{\rm p} = 0.01\,M_{\rm J}$; panel a), the planet exerts only a minor gravitational perturbation. The global morphology remains dominated by wind-driven angular momentum transport, which naturally produces initially axisymmetric rings and gaps in magnetized disks \citep[e.g.,][]{Suriano17,Suriano19,Can24,Hu22,Nolan23,Hsu24}. The embedded planet introduces only localized distortions without significantly altering the large-scale structure. Panel (a) already shows that the large-scale ring-gap morphology remains predominantly controlled by the wind-driven disk dynamics at late times. The accompanying animation further demonstrates that this qualitative behavior persists throughout the evolution. 

For the intermediate-mass planet ($M_{\rm p} = 0.1\,M_{\rm J}$; panel b), the disk enters a transitional regime. The surface density simultaneously exhibits magnetic wind-driven rings and gaps, non-axisymmetric vortices associated with Rossby wave instability, and spiral features excited by the planet. Panel (b) illustrates the coexistence of these structures, including the fragmented arcs and spiral features produced by the interaction between MHD-driven substructure and planetary perturbations. Neither mechanism fully dominates; instead, the morphology reflects the comparable influence of MHD processes and planetary gravitational forcing.

In contrast, the Jupiter-mass planet ($M_{\rm p} = 1\,M_{\rm J}$; panel c) fundamentally reshapes the disk. A deep, wide gap forms rapidly, accompanied by strong, large-scale spiral density waves that initially dominate the global appearance. The snapshot in panel (c) shows the evolved gap-and-spiral morphology produced by the Jupiter-mass planet, including the broad depleted gap and the large-scale spiral shocks extending across the disk. The accompanying animation further illustrates how these planet-driven structures emerge early and subsequently interact with the MHD-generated substructure. At later times, MHD processes increasingly modify the spiral-driven pattern, introducing additional ring–gap substructure within the planet-modified environment.


\subsection{Meridional structure and competing planet- and wind-driven motions}
\label{subsec:gas_vertical_structure}

\begin{figure*}
   \centering
\includegraphics[width=\linewidth]{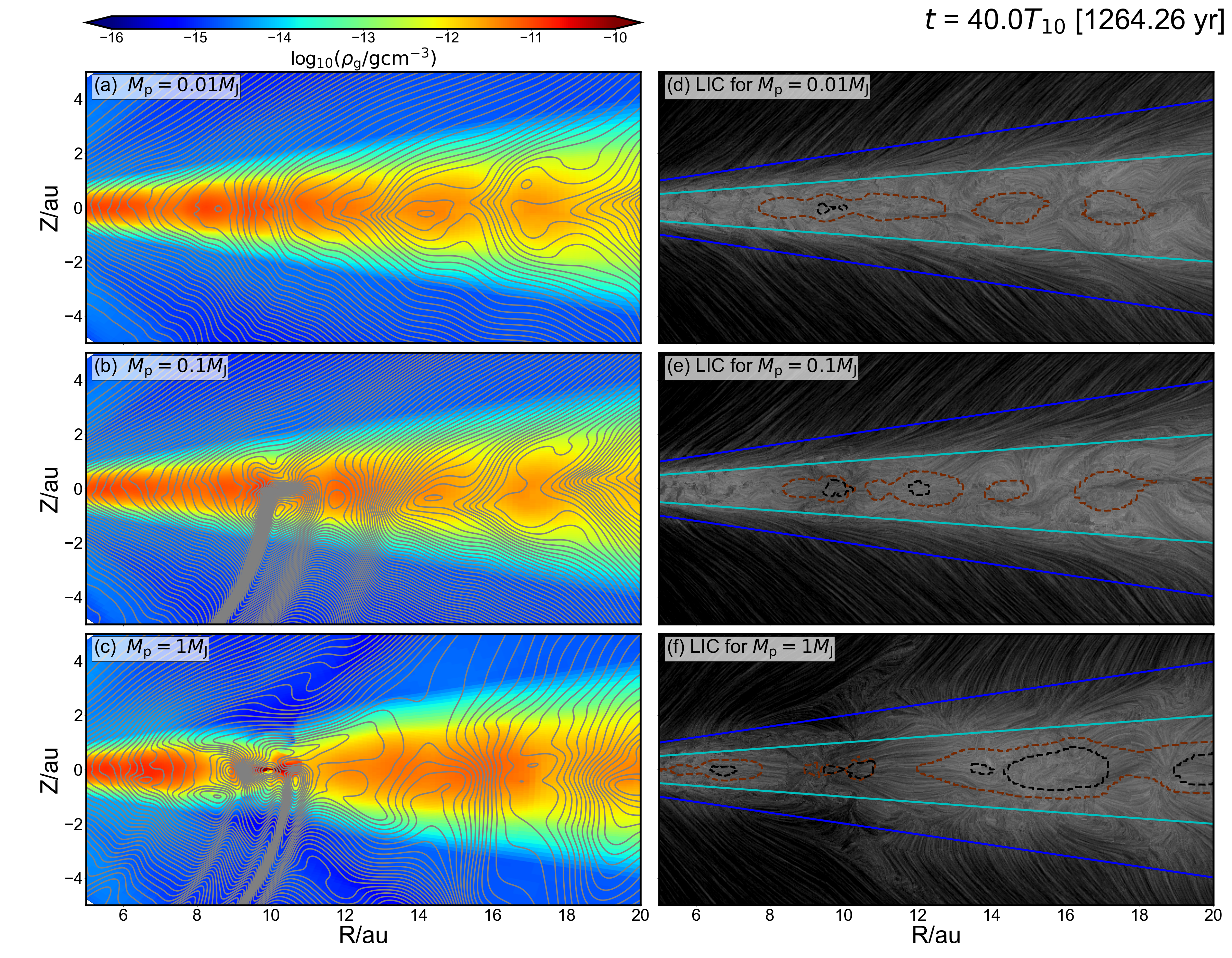}
    \caption{The structures of the meridional plane slices centered at the azimuthal location of the embedded planet ($\phi=\pi$) for Model M001J, M01J, and M1J. Plotted in panels (a)-(c) are the gas density distribution (color map) and magnetic field lines (gray lines). Panels (d)-(f) show LIC (line integral convolution) streamlines for the poloidal gas motions. The dashed lines are isocurves of gas density, respectively 0.6 (brown) and 1.0 (black) times the initial value, highlighting the gas rings. The cyan and blue lines correspond to two and four scale heights, respectively. An animated version of the figure can be found at \url{https://figshare.com/s/073dfca9879e00c4f25e}.
    }
   \label{fig: pure_gas_meridian}
\end{figure*}

We now examine the gas structure in the meridional plane ($R$--$z$) to diagnose how the wind-driven motions (through magnetic stresses) in the meridional plane of a non-ideal MHD disk interact with perturbations introduced by an embedded planet through gravitational torques. Figure~\ref{fig: pure_gas_meridian} shows meridional slices taken through the azimuthal location of the planet ($\phi=\pi$) for Models M001J, M01J, and M1J. Panels (a)--(c) display the gas density overlaid with magnetic field lines, while panels (d)--(f) show the poloidal gas velocity structure using line integral convolution (LIC).

We begin with the lowest-mass case ($M_{\rm p}=0.01\,M_{\rm J}$), which provides a useful reference for the intrinsic wind-driven disk structure. In agreement with previous studies, magnetic field lines initially exhibit the characteristic pinched configuration near the midplane in an inner disk region associated with wind launching, which expands in size over time \citep[e.g.,][]{Hu22,Hsu24}. The midplane current sheet associated with the pinched field lines later breaks up through magnetic reconnection, concentrating the poloidal magnetic flux preferentially in some regions (low-density gaps) while reducing it in others (denser rings), consistent with previous studies of non-ideal MHD disks without embedded planets \citep[e.g.,][]{Wang19,Cui21,Hu22,Hsu24}. 

The corresponding velocity field (panel d) reveals persistent, fast inward accretion streams near the midplane; the associated animation further demonstrates that these flows emerge early in the disk's evolution and remain dynamically critical throughout the development of rings and gaps. The midplane accretion streams dominate the flow pattern in the more strongly magnetized gap regions. They generate vertical shear between them and the layers above and below the midplane in the denser rings, driving large-scale meridional circulation cells. Even at late times, when wind-driven ring--gap structures are fully developed, the global meridional flow pattern remains relatively well organized, with coherent accretion channels and circulation cells. The embedded planet introduces only weak local perturbations and does not significantly disrupt the wind-driven accretion and circulation patterns.

The Jupiter-mass case ($M_{\rm p}=1\,M_{\rm J}$) exhibits a qualitatively different interaction between planetary forcing and wind-driven accretion and circulation. In hydrodynamic disks with $M_{\rm p} \gtrsim M_{\rm th}$ (thermal mass), spiral density waves excited by the planet steepen into shocks that deposit angular momentum locally and drive gap opening \citep{Rafikov2002}. Consistent with this picture, in the meridional plane (panels c and f), the planet produces strong poloidal deflections near its orbit, generating localized overturning flows and vertical motions associated with the spiral shocks. Three-dimensional simulations of gap-opening planets have shown that such spiral shocks can induce significant vertical stirring and meridional circulation (sometimes described as shock bores) in the vicinity of the planet \citep{BoleyDurisen2006,FungChiang2016}. 

These planet-driven motions appear early in the simulation and initially dominate the local velocity structure near the midplane before the wind-driven rings and gaps fully develop. The early-time flow shows clear distortion of the magnetic wind-driven midplane accretion streams, particularly in the vicinity of the planet and at the edges of the planet-opened gap. Because these accretion streams are closely associated with the thin midplane current sheet where the horizontal magnetic field reverses sign, their distortion implies that the current sheet itself is strongly perturbed as well. We return to this point in the next subsection, when examining the radial gas transport, where the current sheet becomes directly visible through the plasma-$\beta$ structure (see Fig.\ref{fig:faceon_flux}b below).

In addition, a well-defined circumplanetary disk (CPD) forms within the Hill sphere of the $1\,M_{\rm J}$ planet (panel c). The CPD is rotationally supported and is threaded by a concentrated bundle of poloidal magnetic field lines, indicating significant magnetic flux accumulation in the immediate vicinity of the planet. This concentration likely results from the advection and compression of magnetic flux by gas entering the Hill sphere and circularizing around the planet. The enhanced magnetic field strength within the CPD perturbs the surrounding magnetic topology, modifying the structure of field lines just outside the Hill sphere. Thus, in the Jupiter-mass case, the planet not only reshapes the gas density and velocity field through spiral shocks, but also locally reorganizes the magnetic flux distribution at scales comparable to its Hill radius. The presence of a magnetized CPD embedded within a wind-launching disk raises interesting questions about how gas and solids are delivered to the planetary vicinity, which we will discuss briefly in \S~\ref{subsec:cpd}.

However, the magnetic wind-driven fast-inward accretion streams are not globally eliminated by the Jupiter-mass planet. Instead, they are distorted and partially rechanneled around the developing planet-opened gap and planet-driven spiral waves. At later times, the flow becomes vertically stratified: near the midplane, the planet strongly reshapes the density and velocity structure, while at higher altitudes, the wind-driven accretion and circulation persist and continue to organize the poloidal flow. Thus, the planet does not replace the wind-driven accretion system; rather, it locally reshapes it within a globally magnetically driven disk.

Magnetic flux redistribution in the disk outside the planet-opened gap further illustrates this interaction. In wind-driven ring--gap systems, magnetic flux typically concentrates in low-density gap regions, producing pinched field-line geometries that support rapid inward accretion. In Model M1J, by contrast, magnetic flux is concentrated not only in gaps but also in localized, non-axisymmetric structures situated within the high-density rings (such as the vortex near $R\sim 16$ au in panel c of Fig.~\ref{fig: pure_gas_meridian}). These flux concentrations coincide with planet-modified structures, including overdense spiral-shock features and associated vortices likely generated through Rossby-wave instability, and do not display the same pinched morphology characteristic of purely wind-generated gaps.

The intermediate-mass case ($M_{\rm p}=0.1\,M_{\rm J}$) exhibits a behavior between these low and high-mass limits. The planet partially distorts the wind-driven meridional accretion and circulation and modifies the local magnetic topology, but it does not fully suppress the fast inward accretion streams characteristic of the wind-launching disk. The resulting flow geometry reflects a competition between planetary perturbations and global MHD-driven motions, consistent with the mixed column-density morphological regime identified in Section~\ref{subsec:gas_morphology}.

\subsection{Radial gas transport and gap permeability}
\label{subsec:gas_transport}

Having established the morphological differences between the wind-dominated and planet-dominated regimes (Section~\ref{subsec:gas_morphology}) and the corresponding meridional flow structures (Section~\ref{subsec:gas_vertical_structure}), we now quantify how these structural differences translate into radial mass transport. In particular, we examine whether the rings and gaps, whether generated by MHD processes or by the embedded planet, act as barriers to inward gas transport.

\begin{figure*}
   \centering
\includegraphics[width=\linewidth]{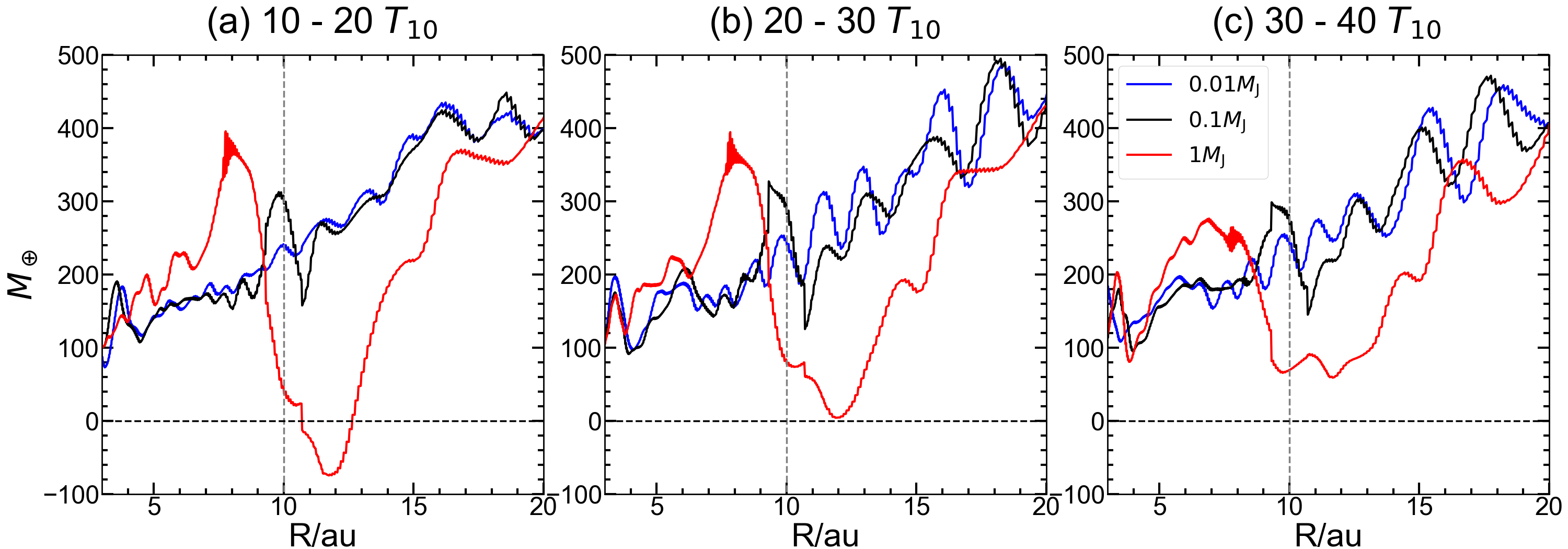}
    \caption{Integrated gas mass (in units of Earth's mass $M_\oplus$)  crossing a given radius $R$ over successive $10\,T_{10}$ intervals for Models M001J ($0.01\,M_{\rm J}$, blue), M01J ($0.1\,M_{\rm J}$, black), and M1J ($1\,M_{\rm J}$, red). Positive values correspond to net inward transport. The vertical dashed line marks the planet's orbit at 10~au. The integral is evaluated within two gas scale heights about the midplane. Wind-driven rings and gaps produce radial modulation in the transport profiles but do not prevent inward mass flux. In contrast, the Jupiter-mass planet initially produces a strong suppression—and even local reversal—of transport just exterior to its orbit, before the gap becomes partially permeable at later times. 
    }
   \label{fig:integrated_mass_transport}
\end{figure*}

Figure~\ref{fig:integrated_mass_transport} shows the cumulative gas mass crossing a given cylindrical radius $R$ over successive $10\,T_{10}$ intervals, where $T_{10}$ is the orbital period at 10~au. Positive values indicate net inward transport, and the integral is taken within two gas scale heights about the midplane. This diagnostic measures whether there is a net amount of gas that actually crosses a given radius over a finite time interval, independent of short-term oscillations in the instantaneous accretion rate.

We first consider the lowest-mass case ($0.01\,M_{\rm J}$), which serves as a benchmark for transport in a wind-launching disk with only weak planetary perturbations. In all time intervals shown in Fig.~\ref{fig:integrated_mass_transport}, the net transport remains positive at essentially all radii, despite the development of pronounced wind-driven rings and gaps in surface density (Section~\ref{subsec:gas_morphology})\footnote{The overall radial increase in the transported mass reflects, at least in part, the larger mass reservoir available at larger radii, which may be specific to the initial conditions adopted in our disk model.}. In particular, there is no persistent suppression of inward transport near the planet orbit at 10~au. 

The radial modulation (``wiggles'') in the transport profiles correlates with the underlying magnetically generated ring--gap structure. These modulations reflect variations in local surface density and magnetic stress across rings and gaps. They modulate the magnitude of inward transport rather than reversing its direction. This behavior is consistent with the meridional structure discussed in Section~\ref{subsec:gas_vertical_structure}, where fast midplane accretion streams persist within both rings and gaps and drive sustained inward mass flux. Thus, in the absence of a sufficiently massive planet, magnetic wind-generated rings and gaps do not constitute dynamical barriers to radial gas transport.

The Jupiter-mass case exhibits qualitatively different behavior. During the early stages of evolution (before $\sim 20\,T_{10}$), the integrated transport profile shows a pronounced suppression of the inward mass flux just outside of the orbit of the planet. In some intervals, a narrow radial region centered near $\sim 12$~au displays net outward radial transport (negative values in Fig.~\ref{fig:integrated_mass_transport}a). At the same time, inward transport interior to the planet orbit is enhanced relative to the low-mass cases.

This pattern reflects the interaction between the wind-driven accretion flow and the torques associated with the planet-driven spiral density waves. In the outer disk, spiral shocks locally deposit angular momentum into the gas, thereby weakening or reversing the underlying inward transport driven by magnetic stresses. As a result, radial transport near the outer gap edge is temporarily suppressed and can even become outward in a narrow radial zone. The location of this transport reversal coincides with the outer edge of the planet-opened gap seen in the surface density maps (Section~\ref{subsec:gas_morphology}).  In this early phase, the planet-driven torques therefore strongly oppose the wind-driven inward transport across the outer gap edge.

At later times ($\gtrsim 30\,T_{10}$), the transport behavior in the $1\,M_{\rm J}$ case changes. The region of previously negative (outward) transport near the outer gap edge becomes weakly positive (inward), indicating that gas can accrete across the planetary orbit when averaged over finite time intervals (Fig.~\ref{fig:integrated_mass_transport}c). Although the inward flux remains reduced relative to neighboring radii, the gap is no longer an effective global barrier to radial transport.

\begin{figure*}
\centering
\includegraphics[width=\linewidth]{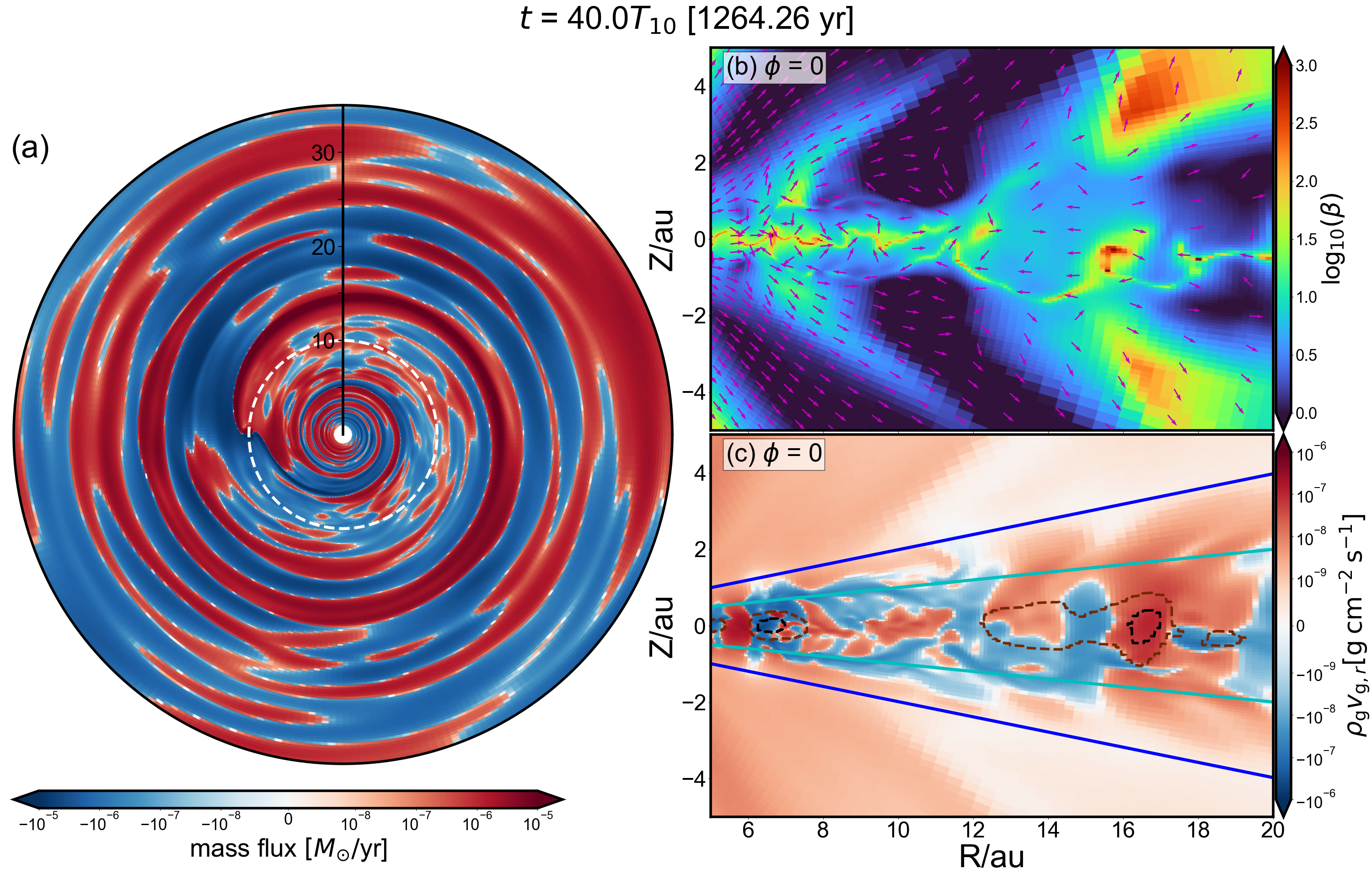}
\caption{Panel (a) shows the face-on map of the vertically integrated radial gas mass flux for the $1\,M_{\rm J}$ model at $t=40\,T_{10}$. Blue indicates inward radial transport and red outward radial transport. 
The white dashed circle corresponds to the planet orbit at $r$ = 10~au. 
Rather than forming an axisymmetric barrier, the outer gap region contains alternating azimuthal sectors of inward and outward radial transport associated with the planet-driven spiral structure. The azimuthally averaged transport measured in Fig.~\ref{fig:integrated_mass_transport} therefore reflects the net balance between these competing sectors. Panel (b) displays the plasma-$\beta$ distribution with gas velocity vectors on the meridional plane with $\phi=0$ (opposite of the planet). The thin high-$\beta$ layer along the disk midplane marks the current sheet where both $B_r$ and $B_\phi$ reverse sign. Panel (c) shows the radial mass flux $\rho v_r$ in the same meridional plane. 
Animations showing the time evolution of the radial mass-flux maps for the Models $0.01\,M_{\rm J}$ at \url{https://figshare.com/s/e4937f2ad29e44cbf782}, $0.1\,M_{\rm J}$ at \url{https://figshare.com/s/3f5fe160c44db761aad7}, and $1\,M_{\rm J}$ at \url{https://figshare.com/s/a1af4e53b106e23fce7b}. 
}
\label{fig:faceon_flux}
\end{figure*}

Figure~\ref{fig:faceon_flux}a shows that the instantaneous flow pattern near the planet-opened gap at $t=40\,T_{10}$ is highly non-axisymmetric. Rather than forming a continuous axisymmetric barrier, the outer gap region contains alternating azimuthal sectors of inward and outward radial transport. The azimuthally averaged transport profiles in Fig.~\ref{fig:integrated_mass_transport} therefore represent the net balance between these competing sectors. The wind-driven accretion flow itself also exhibits strong azimuthal variability, producing patchy arcs and short spirals consisting of alternating sectors of inward and outward radial transport. The accompanying animations further illustrate that these structures evolve significantly with time and intermittently reorganize into spiral-shaped inflow channels that penetrate the gap. In the presence of a massive planet, spiral shocks locally weaken or reverse the inward transport along the spiral arms, while the inter-arm regions remain more weakly perturbed and continue to accrete under the action of magnetic stresses. As a result, inward transport occurs primarily in the inter-arm regions, forming spiral-shaped inflow channels that intermittently penetrate the gap.

Panels (b) and (c) of Fig.~\ref{fig:faceon_flux} further show that the azimuthally intermittent transport pattern is accompanied by a substantial reorganization of the vertical accretion structure. The high-$\beta$ current-sheet layer is no longer confined to a single thin midplane sheet but becomes vertically distorted and fragmented into localized accretion channels above and below the midplane. At early times, the planet interacts with dense gas near the disk midplane, driving a broad and relatively continuous outward expansion exterior to its orbit. However, as the gap deepens, the local gas density decreases, reducing the amount of mass available for the planet to torque; in this sense, the planet's torque-driven outward transport is intrinsically self-limiting. 

In contrast, wind-driven accretion is governed by magnetic stresses arising from the interplay between the magnetic field configuration and the distributions of mass and rotation in both the disk and the wind—particularly differential rotation along field lines—rather than by the local disk density alone. As a result, the accretion flow does not diminish in proportion to the local gas density, and can remain significant—or even locally enhanced in terms of mass flux per unit area—within the low-density gaps relative to dense rings. This behavior is already evident in the mass-flux structures shown in Fig.~\ref{fig:faceon_flux}, while the accompanying animations further demonstrate that the localized accretion channels remain highly time dependent and azimuthally intermittent throughout the evolution.

Consequently, as the planet-driven outward mass flux weakens and becomes increasingly fragmented as the gap deepens, the wind-driven accretion flow does not diminish correspondingly, but instead reorganizes into localized channels associated with a distorted and vertically displaced current-sheet structure traced by high-plasma-$\beta$. The meridional snapshots in panels (b) and (c) show that these channels are no longer confined to a single continuous midplane layer but instead occupy regions above and below the midplane. The accompanying animations further show that these vertically displaced accretion channels evolve significantly with time.

The azimuthally averaged transport profiles in Fig.~\ref{fig:integrated_mass_transport} therefore represent the net outcome of a competition that becomes progressively less co-spatial. At early times, both the planet-driven expansion and the wind-driven accretion primarily act on the same dense midplane gas, allowing the outward transport associated with spiral torques to efficiently suppress inward flow across the outer gap edge. At later times, the weakening of the planet-driven expansion and the redistribution of magnetic-stress-driven accretion into vertically and azimuthally localized channels reduce this direct competition. Inward transport can then proceed through regions where the planet's outward forcing is ineffective, even as other regions continue to expand. 

The $0.1\,M_{\rm J}$ model exhibits behavior intermediate between the two extremes but remains qualitatively wind-dominated. Although a noticeable gap appears in the surface density at late times (Fig.~\ref{fig: pure_gas_face_on_view}b), the integrated transport remains positive at nearly all radii and times (Fig.~\ref{fig:integrated_mass_transport}, black curves). No persistent transport barrier develops near 10~au. This highlights an important conceptual point that the presence of a surface density depression alone does not guarantee stoppage or reversal of radial gas transport; a sufficiently strong planetary torque relative to the wind-driven stress is required to produce a genuine suppression of net radial gas flow.


\subsection{Local suppression and restructuring of disk winds by a massive planet}
\label{subsec:wind_modification}

\begin{figure*}
\centering
\includegraphics[width=0.9\textwidth]{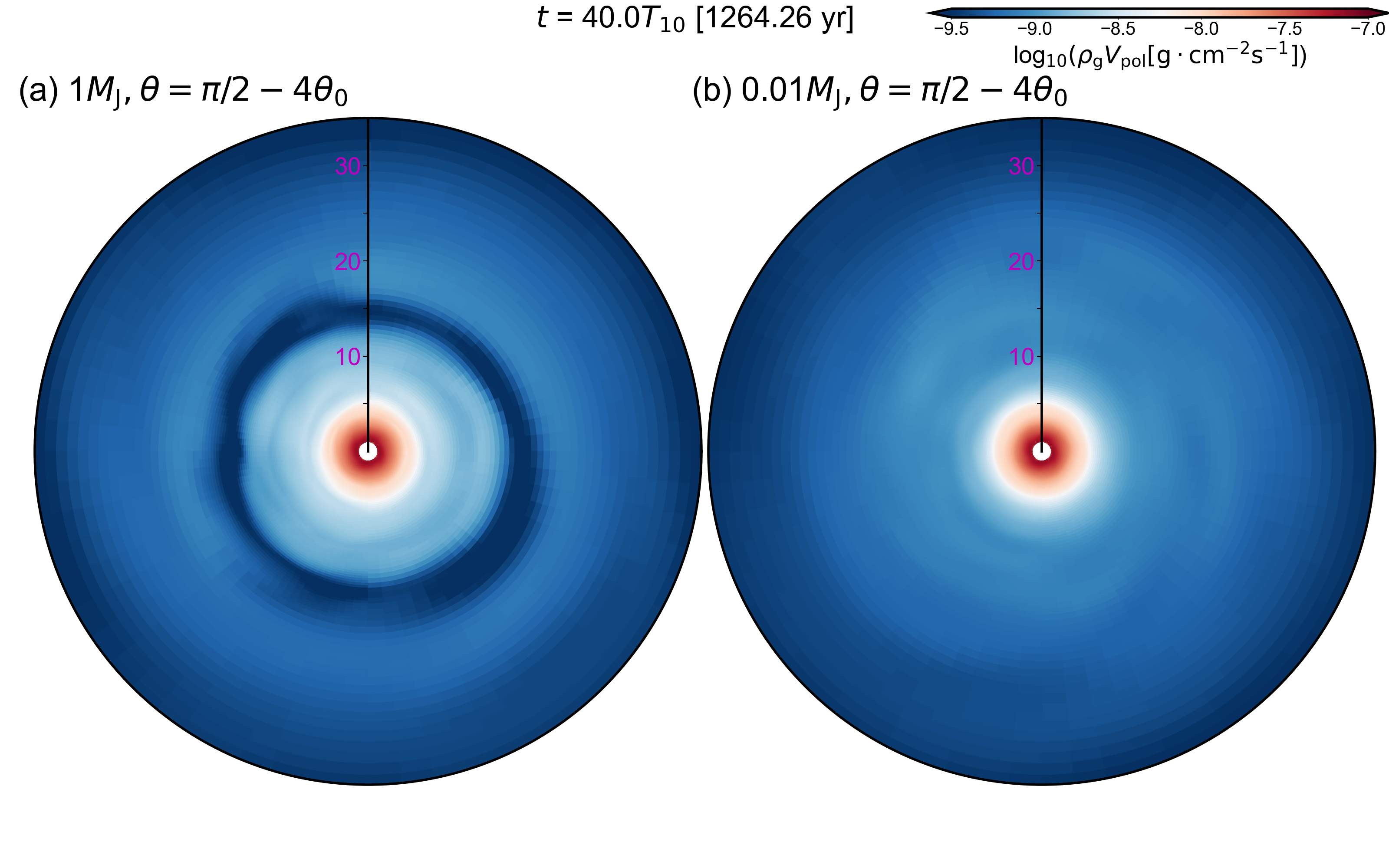}
\caption{
Poloidal mass flux per unit area, $\rho_{\rm g} v_{\rm pol}$, on a cone 8 angular scale heights above the midplane at $t = 40\,T_{10}$ for (a) the $1\,M_{\rm J}$ model and (b) the $0.01\,M_{\rm J}$ model. The color scale is logarithmic, with darker colors indicating lower mass flux. In the low-mass case, the wind-launching structure remains relatively smooth and nearly axisymmetric. In contrast, the Jupiter-mass planet produces a pronounced annular region of suppressed poloidal mass flux coincident with the planet-opened gap. An animation showing the time evolution is available at \url{https://figshare.com/s/124a54720cea985b65d7}.
}
\label{fig:wind_flux}
\end{figure*}

A massive planet affects the structure and gas transport not only in the disk but also in the wind. This is illustrated in Figure~\ref{fig:wind_flux}, which shows the poloidal mass flux per unit area, $\rho v_{\rm pol}$, near the base of the wind at 8 scale heights (or $\sim 23^\circ$) above the disk midplane for both the $1\,M_{\rm J}$ and $0.01\,M_{\rm J}$ models. In the low-mass case, the wind-launching structure remains largely intact, exhibiting a relatively continuous distribution of poloidal mass flux, with moderate spiral-like perturbations reflecting the Rossby-wave-modified, non-axisymmetric rings and gaps in the underlying disk. In contrast, the Jupiter-mass planet produces a progressively stronger reduction of the poloidal mass flux in the vicinity of its orbit, as seen in the animation. At late times ($t \sim 40\,T_{10}$), this reduction manifests itself as a shell-shaped region of suppressed wind mass flux that extends from the gap on the disk surface into the overlying wind, indicating a direct connection between the planet-opened gap and the weakened wind. A similar vertically extended gap structure is also visible in the global non-ideal MHD simulations of \cite{Aoyama23} (see their Fig.~10). 

The strong suppression of the poloidal wind mass flux above the planet-opened gap is closely associated with the substantial gas depletion within the gap, which naturally reduces the amount of material available for wind mass loading. At the same time, the magnetic-field geometry near the gap, particularly in the circumplanetary-disk region, is significantly distorted relative to the low-mass-planet case, suggesting that modifications of the local magnetic structure may also contribute to the altered wind-launching behavior. A quantitative separation of the relative roles of gas depletion and magnetic-field modification is beyond the scope of the present work.

The reduced mass flux in the wind does not imply the disappearance of radial accretion in the disk across this region. Instead, it indicates that the disk accretion process there is no longer governed primarily by the original, relatively smooth wind-braking channel. To evaluate the angular momentum transport mechanism operating across the planet-opened gap, we analyzed the contributions from Maxwell stresses and the wind torque following the framework of \citet{Aoyama23}. We find that the Maxwell stresses dominate the angular momentum transport over most of the gap region, whereas the wind torque becomes increasingly important near the outer gap edge and eventually dominates outside the gap. These results support the interpretation that the continued accretion across the gap is primarily mediated by magnetic stresses associated with the locally distorted magnetic-field structure within the gap, while the large-scale wind torque remains more important outside the gap region.

The gas-flow and magnetic-field structures discussed above broadly confirm the behavior found in previous global non-ideal MHD simulations of planet-hosting, wind-launching disks \citep{Aoyama23,Wafflard23,Hu25}. In the present work, these gas structures provide the physical framework for interpreting the dust transport analyzed in Section~\ref{sec:dust_dynamics_transport_feedback}.

\section{Dust dynamics and transport in magnetized planet-hosting disks}
\label{sec:dust_dynamics_transport_feedback}


We analyze three models that include aerodynamic coupling between dust and gas and contain embedded planets of $0.01$, $0.1$, and $1\,M_{\rm J}$, respectively. Each model includes five dust sizes following an MRN-type size distribution with a power-law index of $-3.5$. The initial total dust-to-gas mass ratio is $\epsilon=0.01$, and the dust is initially distributed within four gas scale heights.

In magnetized wind-launching disks, dust transport is governed by the interplay between aerodynamic drift and radially and vertically structured gas flows produced by magnetic stresses. In the non-planet simulations of \citet{Hsu25}, dust grains settle toward the midplane and drift inward until their migration is slowed or halted by the complex meridional flow structure associated with wind-driven accretion. 
In particular, while magnetic braking removes angular momentum near the disk surface to drive rapid surface accretion, it also magnetically transfers and deposits angular momentum into the disk midplane. This local gain of angular momentum subsequently drives an outward gas flow near the midplane, reducing or even reversing the inward drift of the settled dust layer.
As a result, large grains can accumulate at radii of several au without requiring a conventional pressure maximum. 

The planet-hosting simulations considered here build on this framework and allow us to examine how an embedded planet modifies the dust transport pathways established by the magnetized disk. To establish a reference point, we first consider the lowest-mass planet model, which is only weakly perturbed by the planet--disk interaction and therefore provides a useful baseline for dust dynamics and transport in a magnetized wind-launching disk, even though most of the results in this baseline case will be similar to the non-planet case previously studied by \cite{Hsu25}.

Although the focus of this section is dust transport, interpretation of the dust behavior requires reference to the gas flow established in Section~\S\ref{sec:gas_dynamics_transport}, which we recap where needed.

\subsection{Dust transport in the lowest-mass planet case}
\label{subsec: dust_lowest-mass}

\begin{figure*}
   \centering
\includegraphics[width=\linewidth]{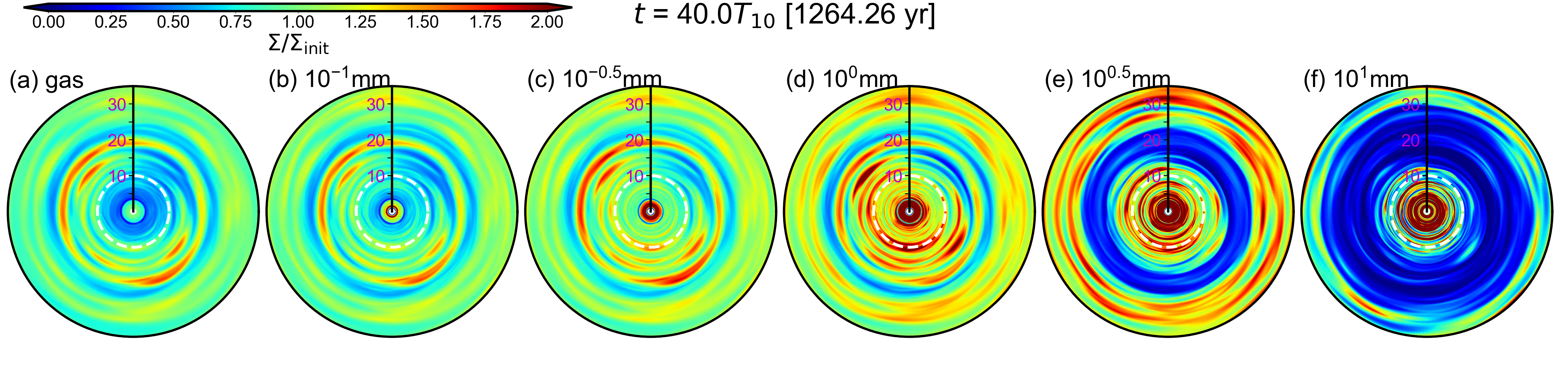}
\caption{
Face-on surface density maps of the gas and different-sized dust grains in Model M001JD10mm01mm5bins at $t=40T_{10}$. All surface densities are normalized to their initial values and shown within a radius of 35 au. The white dashed circles correspond to $r$ = 10~au. The gas develops prominent ring–gap substructures produced by magnetic stresses in the wind-launching disk. Dust grains respond to these structures in a size-dependent manner. Small grains remain well coupled to the gas and largely trace the gas rings and gaps, whereas larger grains become strongly concentrated within the dense rings and/or the inner disk and depleted in the outer disk. The 1~mm and $10^{0.5}$~mm grains are also efficiently concentrated within vortices associated with the Rossby wave instability (RWI). Despite the strong substructure, dust grains can migrate across multiple rings before accumulating at smaller radii, illustrating that the MHD-generated rings and gaps modulate but do not halt radial dust transport. An animated version of this figure is available at \url{https://figshare.com/s/6ec6c472f0b112622b33}.
}
\label{fig: M001_5dust_faceonview}
\end{figure*}

\begin{figure}
   \centering
\includegraphics[width=0.9\linewidth]{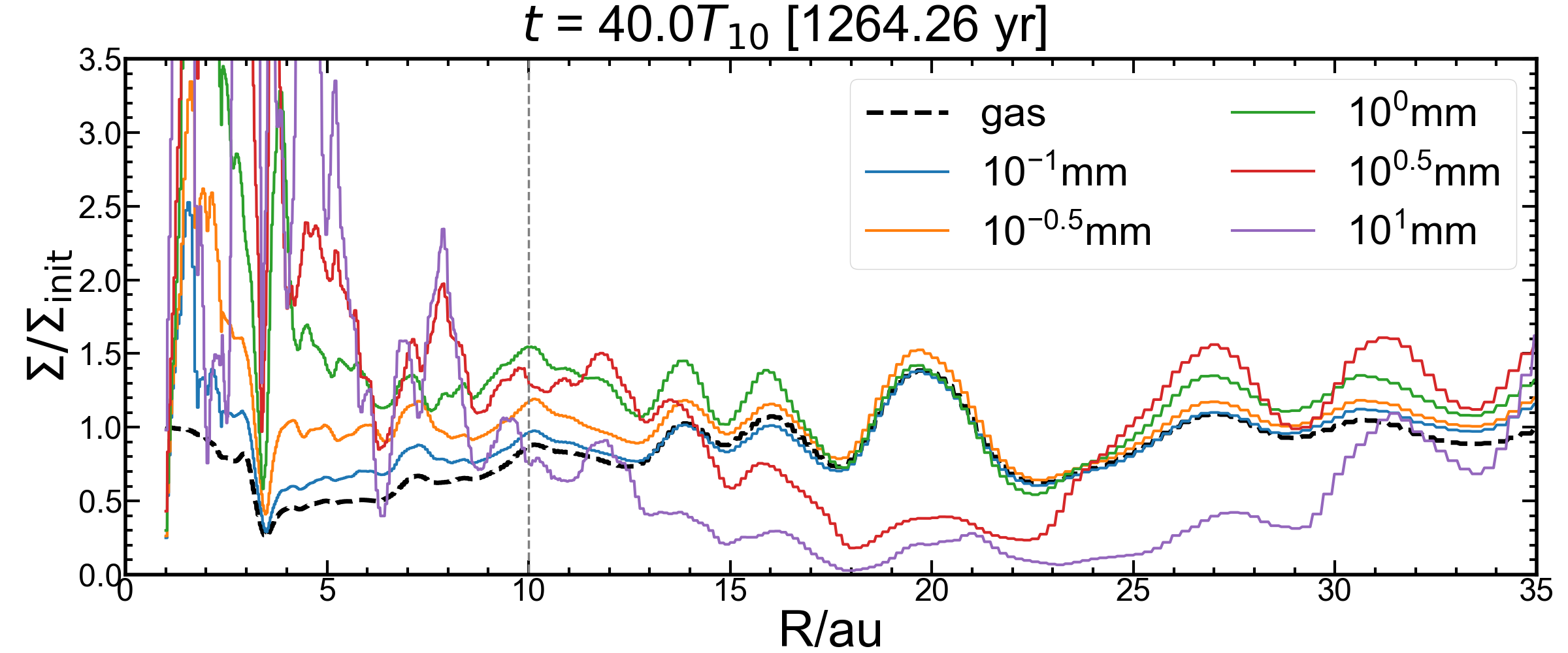}
\caption{
Azimuthally averaged radial surface density profiles of the gas and the five dust populations for Model M001JD10mm01mm5bins at $t=40T_{10}$, normalized to their initial values. The gas distribution exhibits moderate radial variations associated with magnetically generated rings and gaps. Small grains remain well coupled to the gas over most of the disk, whereas larger grains show strong concentration at small radii due to vertical settling and radial drift. Dust surface densities increase sharply inside $\sim$5–10 au, where inward migration slows as a result of the complex meridional gas flow structure produced by wind-driven accretion. An animated version is available at \url{https://figshare.com/s/1b0efb257c05532f310a}.
}
\label{fig: M001_5dust_radial}
\end{figure}

Figure~\ref{fig: M001_5dust_faceonview} shows face-on maps of the gas and dust surface densities at $t=40T_{10}$. The gas exhibits the characteristic ring–gap substructures produced by magnetic stresses in the wind-launching disk, with prominent arcs from the Rossby wave instability. Dust grains broadly follow these structures but show strong size-dependent behavior. Small grains remain tightly coupled to the gas and closely follow its distribution. However, larger grains become increasingly concentrated in the inner disk, while becoming depleted at larger radii (around 20 au). This difference is quantified in Fig.~\ref{fig: M001_5dust_radial}, which shows azimuthally-averaged surface density profiles of the gas and the five dust populations.
The animations accompanying Figs.~\ref{fig: M001_5dust_faceonview} and \ref{fig: M001_5dust_radial} illustrate the time evolution of the dust redistribution process, particularly for large grains.\footnote{We note that the simulations include the mutual drag force between gas and dust and thus incorporate dust-to-gas back reaction. In regions of high dust concentration, particularly near pressure maxima and planet-induced substructures, the local dust-to-gas ratio can increase substantially, potentially enabling dust feedback to modify local gas dynamics and substructure evolution. A more detailed analysis of these feedback effects is beyond the scope of this work; see \citet{Hsu25} for a related discussion in non-planet simulations.}


\begin{figure*}
   \centering
\includegraphics[width=\linewidth]{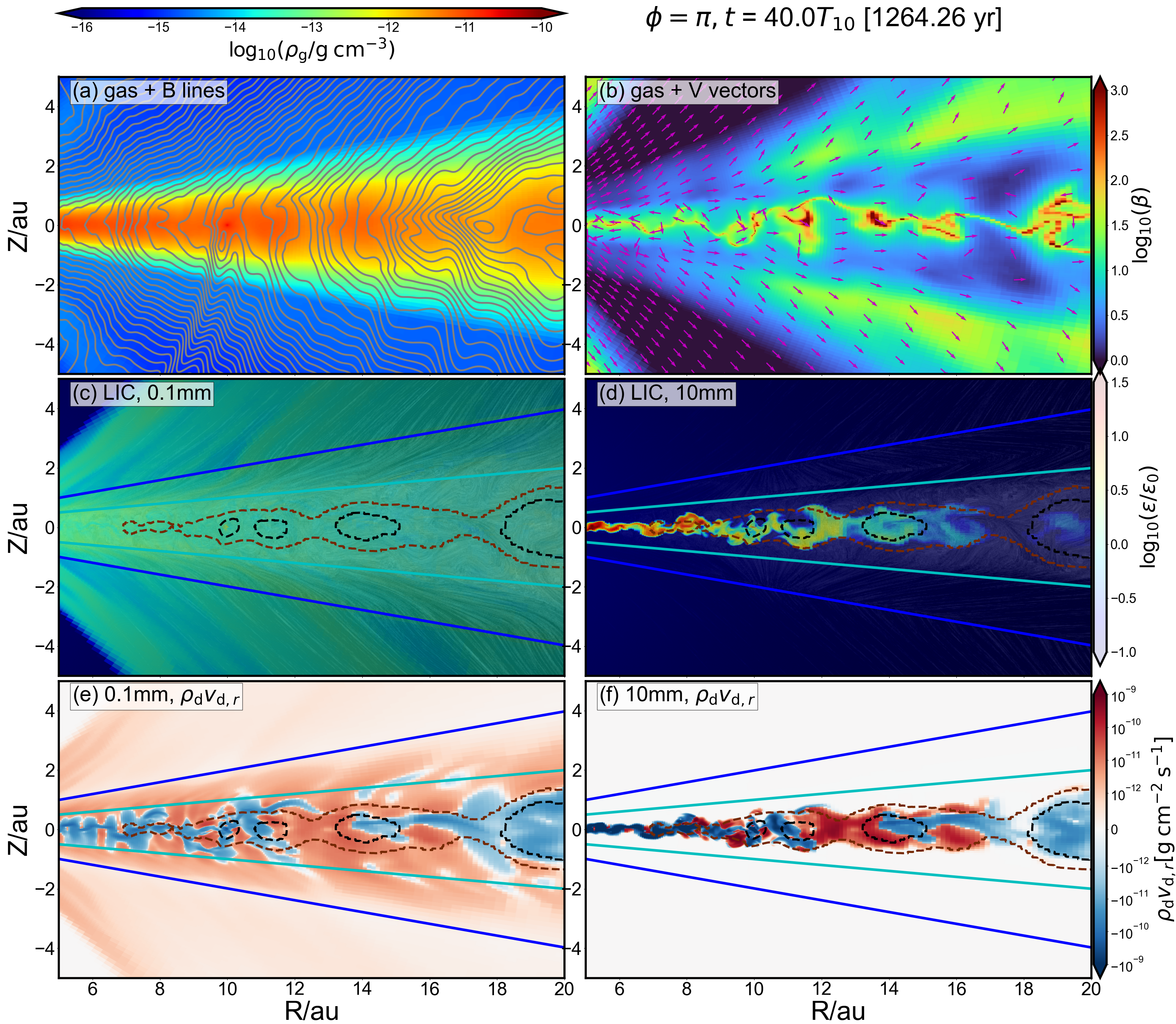}
\caption{
Meridional slices at $\phi=\pi$ (passing through the planet) for Model M001JD10mm01mm5bins illustrating the gas and dust dynamics. 
Panel (a) shows the gas density with magnetic field lines, while panel (b) displays the plasma-$\beta$ distribution with gas velocity vectors. The thin high-$\beta$ layer along the disk midplane marks the current sheet where both $B_r$ and $B_\phi$ reverse sign. 
Panels (c) and (d) show the dust-to-gas ratio $\epsilon$ for representative small (0.1\,mm) and large (10\,mm) grains, with their velocity fields traced by LIC. Panels (e) and (f) show the corresponding radial dust mass flux per unit area $\rho_{\rm d} v_{{\rm d},r}$ in the meridional plane. Small grains remain vertically extended and participate in both midplane and surface transport channels, whereas large grains rapidly settle toward the midplane and migrate primarily through a thin layer near the current sheet, particularly at early times, which can be seen more clearly in the animated version of the figure available at \url{https://figshare.com/s/aafd71e17a5fe4ed4956}.
}
\label{fig: M001_5dust_LIC}
\end{figure*}

The physical origin of the size-dependent dust transport seen in Figs.~\ref{fig: M001_5dust_faceonview}--\ref{fig: M001_5dust_radial} can be understood from the meridional structure of the disk shown in Fig.~\ref{fig: M001_5dust_LIC} at $t=40T_{10}$. In particular, panel (b) highlights a thin high-$\beta$ layer along the midplane that traces the current sheet where both $B_r$ and $B_\phi$ reverse sign. As discussed in Section~\ref{sec:gas_dynamics_transport}, this current sheet is associated with a narrow midplane accretion stream that develops at early times and provides an efficient pathway for radial transport.

Dust responds strongly to this flow structure in a size-dependent manner. Small grains remain well coupled to the gas and therefore populate a vertically extended layer (panel c), participating in both midplane and off-midplane flows (panel e). In contrast, large grains settle rapidly toward the midplane and become concentrated within the thin current-sheet layer (panel d). Because the early-time midplane accretion stream is confined to this same layer, the settled large grains are transported inward particularly efficiently along the midplane, leading to the enhanced local dust-to-gas mass ratio at small radii (see the red regions in panel d) and the prominent depletion of the two largest grains near $\sim$20\,au. The accompanying animation further confirms that the radial transport of the largest grains is initially dominated by fast inward motion near the midplane. As the disk evolves, the flow pattern in the inner disk becomes more complex. Consistent with \cite{Hsu25}, the grains delivered to small radii at early times encounter locations where the midplane gas flow slows or even reverses, allowing them to accumulate and effectively ``park'' there.



\subsection{Dust transport in the most massive planet case}
\label{subsec: dust_massive-planet}

\begin{figure*}
\centering
\includegraphics[width=\textwidth]{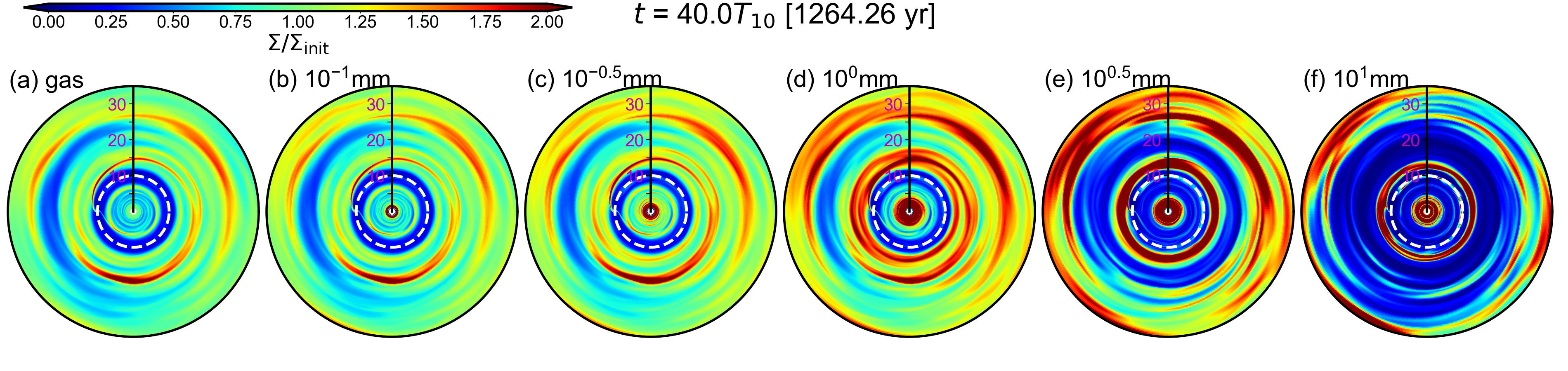}
\caption{
Face-on surface density maps of the gas and different-sized dust grains in the $1\,M_{\rm J}$ planet model at
$t=40\,T_{10}$. All surface densities are normalized to their
initial values and shown within a radius of $35$~au. The white dashed circles correspond to $r$ = 10~au.
The gas exhibits both magnetically generated ring–gap structures and spiral density waves launched by the planet.
Dust responds to these structures in a size-dependent manner.
Small grains remain well coupled to the gas and broadly trace the gas morphology, whereas larger grains exhibit increasingly distinct distributions.
In particular, the largest grains become concentrated both at small radii within the planet-induced gap and in a ring near the gap's outer edge, with strong depletion on both sides of the ring.
These structures arise from the combined effects of magnetic disk dynamics, aerodynamic drift, and planetary torques. An animated
version spanning multiple times is available at \url{https://figshare.com/s/ee0e70bd4f792f1a3822}.
}
\label{fig:1MJ_faceon}
\end{figure*}

\begin{figure}
\centering
\includegraphics[width=\columnwidth]{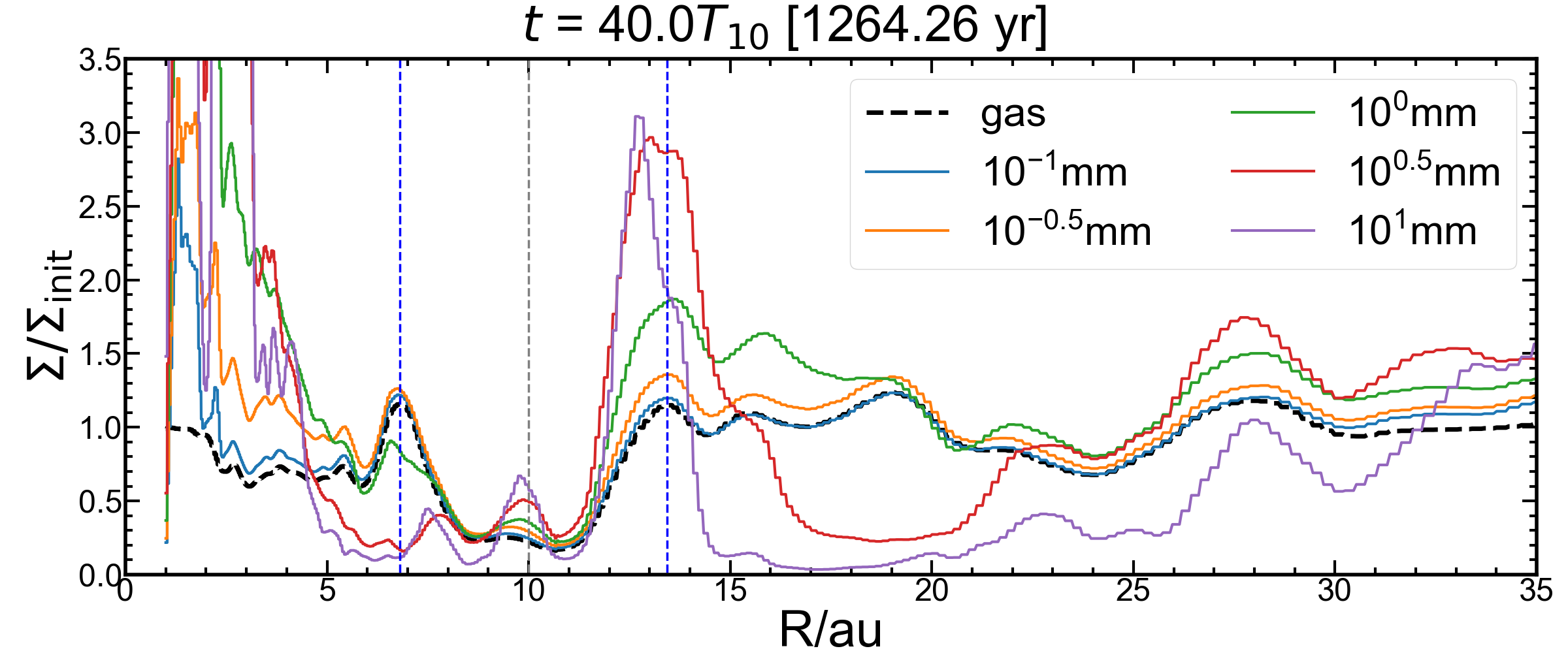}
\caption{
Azimuthally averaged radial surface density profiles of the gas and five dust populations in the $1\,M_{\rm J}$ model at
$t=40\,T_{10}$, normalized to their initial values.
The gas surface density shows a deep planet-induced gap near
$R\sim10$~au together with surrounding ring–gap structures.
Small grains remain well coupled to the gas and therefore broadly follow the gas distribution across most of the disk.
Larger grains preferentially accumulate exterior to the planetary orbit, forming a
prominent dust ring, with strong depletion in the gap and outer disk.
Comparison with earlier snapshots shows that this ring develops
gradually as the planet gap deepens, reflecting the increasing
ability of the outer gap edge to halt the inward drift of large
grains. An animated
version spanning multiple times is available at \url{https://figshare.com/s/d16d3e1f0bbf32d660f1}. 
}
\label{fig:1MJ_sigma}
\end{figure}

\begin{figure*}
\centering
\includegraphics[width=\textwidth]{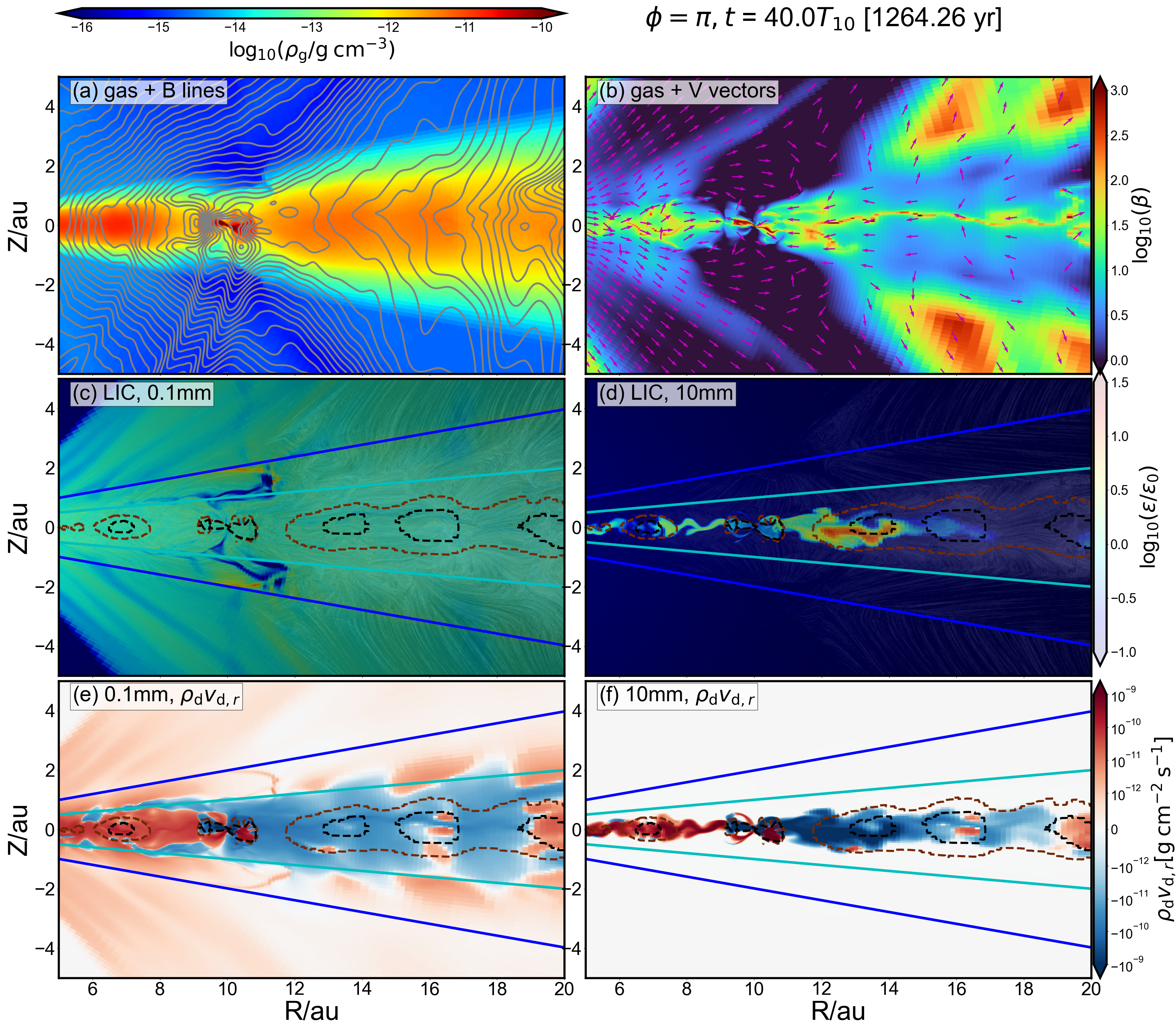}
\caption{
Meridional slices of the $1\,M_{\rm J}$ model at $t=40\,T_{10}$ showing the gas density, plasma-$\beta$, dust-to-gas ratio, and radial dust mass flux per unit area for representative grain sizes.
The thin high-$\beta$ layer along the midplane traces the magnetic current sheet where the horizontal magnetic field components reverse sign and where the fast midplane accretion stream is located.
Compared with the lowest-mass ($0.01\,M_{\rm J}$) planet case, the current sheet and associated accretion stream become strongly distorted by the spiral density waves launched by the massive planet.
This distortion modifies the radial transport pathways of the settled large grains.  An animated version is available at \url{https://figshare.com/s/c278ecfc032813c5c8a4}.
}
\label{fig:1MJ_meridional}
\end{figure*}

We now examine how a massive planet modifies the dust transport pathways established by the magnetized wind-launching disk. Figure~\ref{fig:1MJ_faceon} shows face-on maps of the gas and dust surface densities in the $1\,M_{\rm J}$ model at $t=40\,T_{10}$.
At large radii, the overall morphology remains qualitatively similar to that in the lowest-mass planet case.
The gas continues to exhibit multiple ring–gap substructures associated with magnetic stresses in the wind-launching disk, and the smallest grains remain tightly coupled to the gas and broadly trace these structures.
Progressively larger grains exhibit increasingly distinct distributions from the gas, with stronger concentration in some regions and stronger depletion in others. In particular, the largest grains ($10^{0.5}$ and $10$~mm) are strongly depleted near $\sim20$~au, well outside the orbit of the planet, similar to the $0.01\,M_{\rm J}$ case (see Fig.~\ref{fig: M001_5dust_faceonview}e,f).
As discussed earlier, this depletion results from the early settling of large grains toward the midplane, where they are transported inward efficiently by the fast, wind-driven midplane accretion stream.

Despite this similarity, the presence of a massive planet significantly modifies the global dust distribution.
Strong spiral density waves launched by the planet reshape the gas flow and progressively carve a deep gap near the planetary orbit at $r\sim10$~au.
The largest grains ($a\sim3$--10~mm) respond most strongly to this forcing by the planet's torque. 
Rather than simply concentrating in the inner disk as in the $0.01\,M_{\rm J}$ case, they also accumulate into a prominent ring exterior to the planetary orbit, while becoming depleted both inside the gap and in the outer disk.
The resulting morphology therefore reflects the combined influence of magnetically driven disk dynamics and planetary torques.

The radial redistribution of gas and dust is quantified in Figure~\ref{fig:1MJ_sigma}, which shows azimuthally averaged surface density profiles at $t=40\,T_{10}$.
The gas surface density exhibits a deep gap centered near the planetary orbit, together with surrounding ring–gap structures that retain the imprint of the magnetically generated substructure.
Interestingly, the inner disk still retains a significant population of large grains despite the presence of a deep planet-opened gap. This partly reflects their substantial inward migration during the establishment of the dust-filtering barrier associated with the developing gap, before the super-Keplerian gas motion near its outer edge (where the gas pressure increases rapidly outward) became sufficiently strong to halt their inward migration. We term this interval the ``pre-loading'' phase of large grains. The resulting dust distribution therefore reflects a competition between two processes: rapid inward transport driven by the midplane accretion streams already present in the magnetized disk, and progressively stronger filtering imposed by the planet as the gap deepens.

To better understand how these transport pathways are modified by the massive planet, we examined the meridional structure of the disk.
Figure~\ref{fig:1MJ_meridional} shows meridional slices at the representative late-time epoch $t=40\,T_{10}$. Recall that, in the lowest-mass planet model, the disk maintains a relatively coherent midplane accretion stream associated with the magnetic current sheet at early times, which appears as a thin high-$\beta$ layer and persists for much of the simulation until the full development of magnetic wind-driven rings and gaps (see Fig.\ref{fig: M001_5dust_LIC}b and its animated version). 
In contrast, the $1\,M_{\rm J}$ model exhibits a qualitatively different behavior.
As strong spiral density waves develop, the midplane accretion stream becomes increasingly distorted, particularly near the planetary orbit and the edges of the forming gap. Fig.~\ref{fig:1MJ_meridional}b shows that the thin high-$\beta$ layer associated with the current sheet becomes warped and fragmented as the spiral waves interact with the wind-driven magnetic configuration. This distortion indicates that planet-driven perturbations not only redistribute gas angular momentum via spiral shocks but also modify the magnetic structure that organizes midplane accretion. It affects the transport of the largest grains most strongly, as they settle most quickly and are thus more susceptible to distortion  of midplane accretion. 

A feature of the $1\,M_{\rm J}$ model worth stressing is the presence of two qualitatively distinct dust gaps for large grains, which are most pronounced in 3~mm grains. One gap is associated with the planet at $\sim10$~au, while a second, broader gap appears farther out at $\sim20$~au (see Fig.~\ref{fig:1MJ_faceon}e). These two gaps differ markedly in morphology. The outer gap is typically smoother and more axisymmetric, consistent with dust depletion driven by radial transport. In contrast, the planet-associated gap contains a partial, clumpy, dust ring within the gap region, which is seen more clearly in the larger, 10~mm grains (Fig.~\ref{fig:1MJ_faceon}f). It is also surrounded by a more pronounced dust ring at the outer gap edge. The partial ring is likely the dust trapped by the gas in the planet’s co-orbital (horseshoe) region, modulated by azimuthally non-uniform MHD processes operating inside the gap. The coexistence of these two distinct types of gaps highlights the fundamentally different roles of planetary torques and wind-driven transport in shaping the radial distribution of solids. These morphological differences may allow these two different types of dust gaps to be distinguished observationally through high-resolution ALMA dust continuum observations.

\begin{figure*}
\centering
\includegraphics[width=\linewidth]{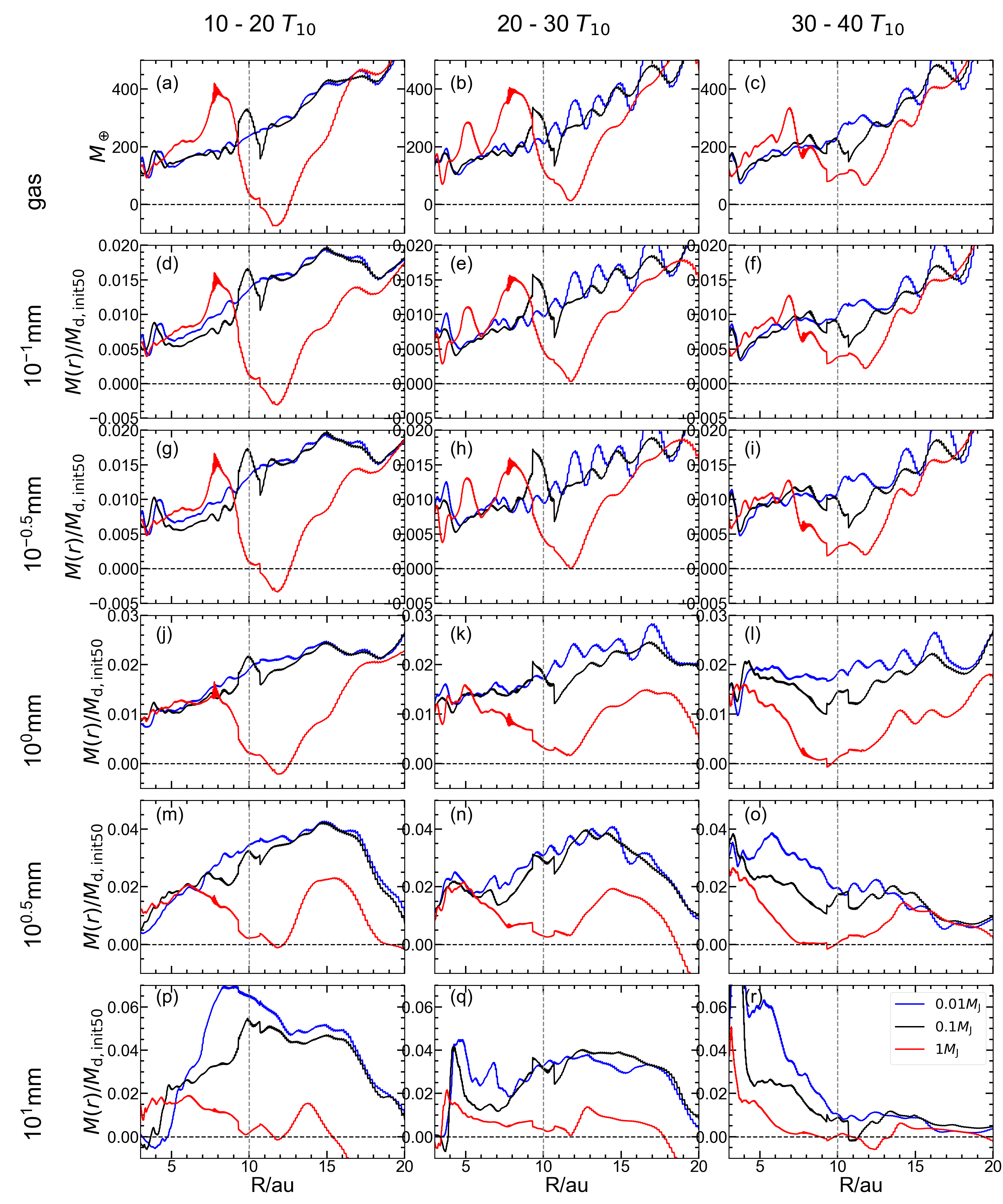}
\caption{Integrated mass crossing a given radius $R$ over three time intervals ($10$--$20$, $20$--$30$, and $30$--$40\,T_{10}$) for gas (in units of Earth's mass) and dust grains of different sizes (in units of their initial mass within the disk in $R=50$~au). Rows correspond to gas and dust sizes of $10^{-1}$, $10^{-0.5}$, $10^{0}$, $10^{0.5}$, and $10^{1}$ mm. Colors denote planet mass: $0.01\,M_{\rm J}$ (blue), $0.1\,M_{\rm J}$ (black), and $1\,M_{\rm J}$ (red). Positive values indicate net inward transport. The vertical dashed line marks the planet's orbit at 10~au. The integral is evaluated within two gas scale heights about the midplane. The weak-planet models ($0.01$ and $0.1\,M_{\rm J}$) exhibit sustained inward transport of gas and small grains across most radii, while the Jupiter-mass model shows a temporary suppression of transport near the outer gap edge. Large grains display additional depletion effects associated with earlier rapid inward migration. }
\label{fig:dust_transport_integrated}
\end{figure*}

\subsection{Comparison of Dust Transport for Different Planet Masses}

To more directly quantify the extent to which planets of different masses act as barriers to radial transport, we examine the integrated mass crossing a given radius $R$ over successive $10\,T_{10}$ intervals. The transport curves in Fig.~\ref{fig:dust_transport_integrated} reveal that the lower planet-mass cases of $0.01\,M_{\rm J}$ and $0.1\,M_{\rm J}$ exhibit very similar behaviors (compare the black and blue curves in the Figure). In particular, the integrated mass crossing profiles for the gas remain broadly positive across most radii and time intervals (blue and black lines in the top panels), indicating sustained inward transport through the magnetically structured disk, consistent with the gas-only simulations discussed earlier (Section~\ref{subsec:gas_transport}). 
A similar pattern is seen for the smaller dust grains ($10^{-1}$ and $10^{-0.5}$~mm; the second and third rows, respectively), whose transport curves closely track those of the gas. Even the $1$~mm grains exhibit broadly similar behavior, with inward transport persisting across most radii. These results indicate that the magnetically generated substructures in the $0.01\,M_{\rm J}$ and $0.1\,M_{\rm J}$ disks do not produce a robust transport barrier for gas or for dust over a wide range of grain sizes.

The largest grains ($10^{0.5}$ and $10$~mm) display a somewhat different temporal behavior. In the early intervals, these grains undergo strong inward transport across all radii, consistent with their rapid settling toward the midplane and subsequent migration along the fast midplane accretion stream identified in the meridional-flow analysis. However, in the final interval ($30$--$40\,T_{10}$), the integrated mass crossing near $\sim20$~au becomes very small for the largest grains in both the $0.01\,M_{\rm J}$ and $0.1\,M_{\rm J}$ cases. This behavior is most naturally explained by the severe depletion of the large-grain reservoir at these radii caused by earlier efficient inward transport, as already evident in the face-on dust maps (Fig.~\ref{fig:1MJ_faceon}). Importantly, the same region continues to allow inward transport of gas and smaller grains, indicating that the inefficient late-time transport of the largest grains does not reflect the formation of a transport barrier.

Taken together, these diagnostics show that the $0.01\,M_{\rm J}$ and $0.1\,M_{\rm J}$ models occupy a common transport regime in which the wind-driven disk dynamics dominate the radial transport of both gas and solids. The magnetically generated rings and gaps modulate the flow and reshape the dust distribution, but do not produce a sustained barrier to inward transport.

The behavior changes more substantially in the $1\,M_{\rm J}$ model. As discussed in Section~\ref{subsec:gas_transport}, the development of a deep planetary gap temporarily suppresses the wind-driven inward gas flow near the outer edge of the gap. This effect is visible in the gas transport curves as a region of strongly reduced or even outward radial transport near $\sim12$~au during the early time intervals. Smaller dust grains largely inherit this behavior because they remain dynamically well coupled to the gas. Their transport continues across the gap at later times, but with a reduced net inward flux compared to the lower planet-mass cases. 

For larger grains, the planetary gap has a more substantial impact. Because these grains experience stronger radial drift and weaker coupling to the gas, they are more sensitive to the pressure structure associated with the planetary gap. In the $1\,M_{\rm J}$ model, the outer gap edge acts as a partial filter for these grains, leading to a reduction in their inward transport relative to the gas and smaller dust particles. Consequently, larger grains accumulate within the region between the super-Keplerian flow (where the gas pressure increases radially outward) and the outer edge of the planet-induced gap (see the red and purple profiles near 12 au in Fig. \ref{fig:1MJ_sigma}). Nevertheless, even in this case, the transport is not completely stopped, and the largest grains can still cross the gap intermittently (see, e.g., panel [n] for the 3~mm grains).

\begin{figure*}
\centering
\includegraphics[width=\linewidth]{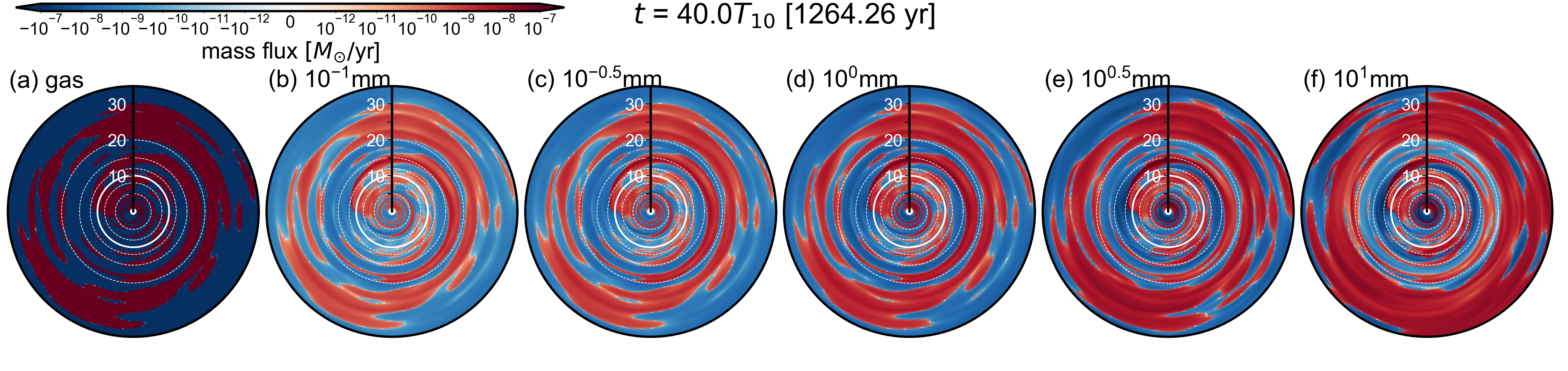}
\caption{Face-on map of the vertically integrated radial mass flux of gas and dust in the $1\,M_{\rm J}$ model at $t=40\,T_{10}$. Blue indicates inward radial transport and red outward radial transport. 
The thick white solid circles correspond to the planet orbit at $r$ = 10~au. The thinner white dashed circles correspond to the $r$ = 5, 8, 12, 15, and 20~au.
The instantaneous transport pattern is highly non-axisymmetric, consisting of alternating azimuthal sectors of inward and outward radial mass flux. These structures arise from the interaction between the wind-driven accretion flow and the planet-driven spiral perturbations. Inward transport occurs primarily in inter-arm regions where magnetic stresses dominate, forming spiral-shaped accretion channels that intermittently penetrate the gap. An animation showing the time evolution of these radial mass-flux maps is available online at \url{https://figshare.com/s/91dae557401f00fe963f} for the 1 $M_{\rm J}$ case, \url{https://figshare.com/s/e63760db2e861c331266} for the 0.1 $M_{\rm J}$ case, and \url{https://figshare.com/s/e1a3165d1e4b48644908} for the 0.01 $M_{\rm J}$ case.}
\label{fig:dust_transport_faceon}
\end{figure*}

Instantaneous flow patterns provide additional insight into this intermittent behavior. Figure~\ref{fig:dust_transport_faceon} shows that the radial transport pattern near the planet-opened gap at $t=40\,T_{10}$ is highly non-axisymmetric. Rather than forming a continuous axisymmetric transport barrier, the outer gap region is organized into alternating azimuthal sectors of inward and outward radial mass flux. These structures arise from the interaction between the wind-driven accretion flow and the planet-driven spiral perturbations. The accompanying animations further demonstrate that these sectors evolve strongly with time and intermittently reorganize into localized inflow channels that penetrate the gap. This strong spatial and temporal intermittency helps explain the behavior of the azimuthally averaged transport profiles.

The azimuthally averaged transport profiles in Fig.~\ref{fig:dust_transport_integrated} therefore represent the net outcome of this spatially structured transport. When spiral-arm regions locally weaken or reverse the wind-driven inflow, the net transport across the outer gap edge can become small or temporarily outward. In the inter-arm regions, however, magnetic stresses continue to drive inward motion. The planetary gap, therefore, behaves as a porous barrier through which gas and dust can continue to cross via localized accretion channels. For the largest grains ($\sim$1–10 mm), the net radial transport across the Jupiter-mass planet gap becomes strongly suppressed at late times, with the azimuthally averaged and time-integrated mass flux often approaching zero (Fig.~\ref{fig:dust_transport_integrated}), although some intervals still show modest net inward transport.
%


\subsection{Dust-to-gas surface density ratio of the circumplanetary disk}
\label{subsec:cpd}

\begin{figure*}
   \centering
\includegraphics[width=\linewidth]{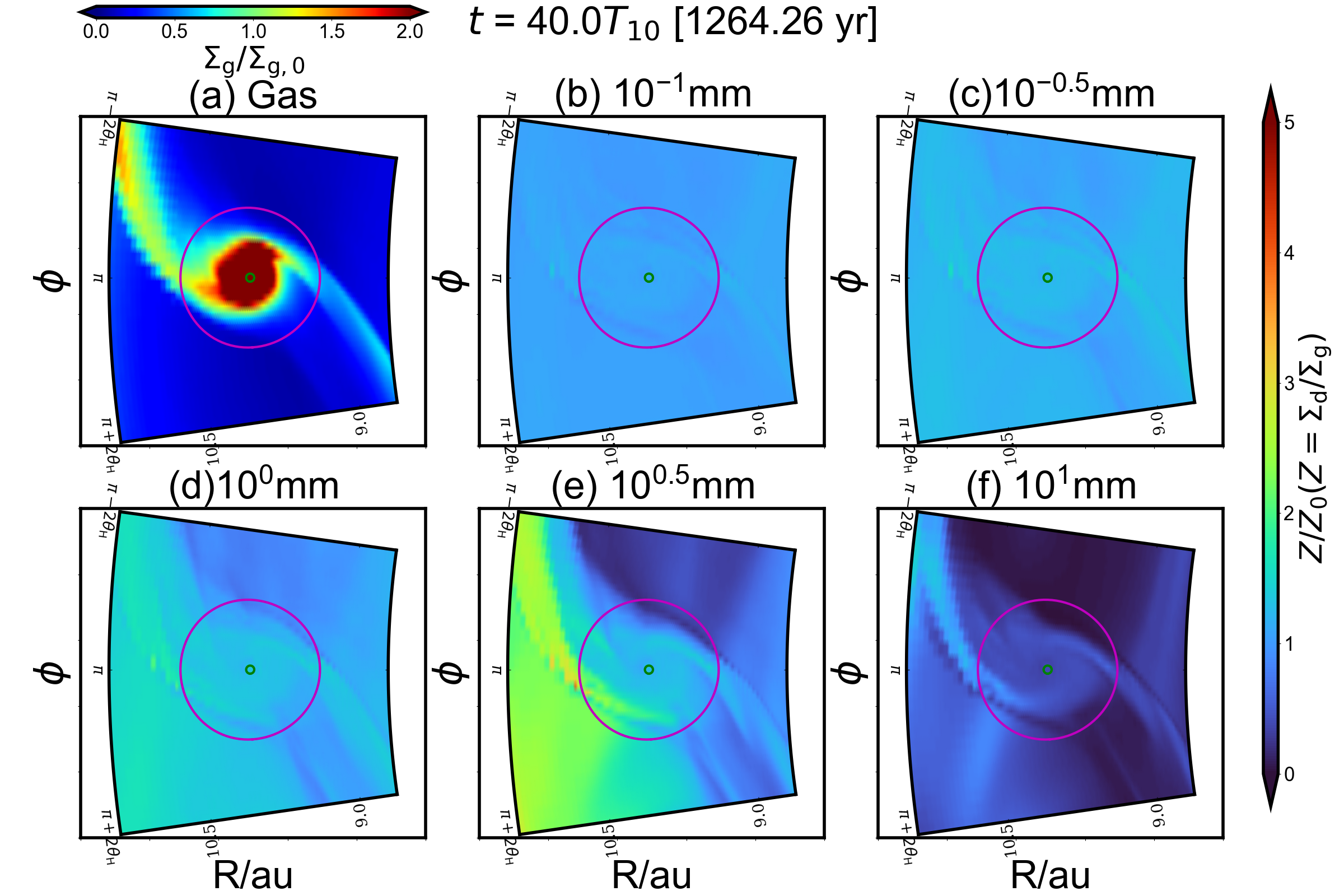}
\caption{Gas and dust spatial distributions near the CPD at $t=40\ T_{10}$. Panel (a) shows the gas surface density normalized by its initial (unperturbed) value, highlighting the dense CPD within the planet's Hill sphere. Panels (b)--(f) show the dust-to-gas surface density ratio for each of the five dust bins, normalized to their initial values. The purple and green circular lines correspond to the planet's Hill Sphere and the softening regions. Significant depletion of the largest grains (particularly the 10~mm grains in panel [f]) is evident. An animated version is available at \url{https://figshare.com/s/152eb665efe1f3f6a345}.
    }
   \label{fig: CPD_Z_faceonview}
\end{figure*}

The planet-carved gap and the co-orbital region in the $1\,M_{\rm J}$ model exhibit a strong, size-dependent dust structure and time-dependent transport. Within this complex environment, a notable feature is the formation of a circumplanetary disk (CPD) within the planet's Hill sphere. Although CPD is not the main focus of this work, our global simulations with dust enable us to examine an important property of CPD: its dust-to-gas mass ratio (i.e., metallicity $Z$), which is relevant to the formation of circumplanetary satellites, such as Galilean moons. Recent multi-frequency observations of the circumplanetary environment around PDS 70c have raised the possibility that some CPDs may be strongly depleted in dust, although the inferred depletion level depends sensitively on the interpretation of the observed radio emission, which is still uncertain \citep{Dom25}. Motivated in part by these observations, we examine the evolution of the dust-to-gas ratio within the CPD in our simulations. We find that the largest grains can become substantially depleted relative to their initial abundance, although the depletion in our simulations is much less extreme than the values inferred for PDS 70c.


Figure~\ref{fig: CPD_Z_faceonview} and its animated time sequence show the face-on distributions of gas surface density and dust-to-gas surface density ratio in the vicinity of the planet. The smallest grains (0.1 and 0.3~mm) closely trace the gas within the CPD, indicating strong coupling to the flow. In contrast, the largest grains (3 and 10~mm) exhibit pronounced spatial and temporal variability. In particular, the dust-to-gas surface density ratio of the 10~mm grains in the inner CPD drops to as low as $\sim 20\%$ of its initial value around $t=35\ T_{10}$, before partially recovering to $\sim 40\%$. The ratio is systematically higher near the outer edge of the CPD, where the flow appears to be more directly connected to the spiral arms outside the Hill sphere, in which large grains are less strongly depleted.

The strong depletion of the largest grains in the inner CPD suggests that the supply of such particles is inefficient. In particular, if the material entering the CPD is delivered primarily through vertically extended and time-dependent flow channels, it may be relatively depleted in large, strongly settled grains, which are more difficult to entrain into these flows. A detailed analysis of dust supply and loss processes within the CPD is beyond the scope of this work.

\section{Discussion}
\label{sec:discussion}

Before discussing specific implications, we note that previous hydrodynamic studies have shown that planet-opened gaps are not perfectly isolating, with both gas and dust able to cross gaps at reduced rates and dust filtration being intrinsically incomplete and size dependent \citep[e.g.,][]{Lubow05,Bitsch18,Huang25c}. Our results extend this picture to non-ideal MHD wind-launching disks, demonstrating that gap permeability can arise not only from the oft-prescribed turbulent diffusion in hydro studies but also from magnetically structured, time-dependent, and non-axisymmetric accretion flows. 

\subsection{JWST observations of inner-disk volatiles}
\label{subsec:jwst_volatiles}

Our results on gas and, especially, dust transport have implications for interpreting recent JWST observations of volatile emission from the inner regions of protoplanetary disks. Early JWST studies have indicated a connection between the outer-disk structure and the volatile content of the inner disk. In particular, \citet{Banzatti2023} found that compact disks exhibit stronger emission from relatively cool H$_2$O compared to disks with extended outer radii and multiple gaps, suggesting that inward drift of icy solids and their subsequent sublimation near the snowline contribute significantly to the inner-disk water reservoir. Subsequent analyses within the MINDS program \citep{Gasman2025} further reported that disks with large outer dust gaps tend to show reduced inner-disk molecular emission, consistent with the idea that pressure bumps can regulate, or partially inhibit, the inward flux of volatile-rich solids.

However, the emerging JWST picture is more complex than a simple dichotomy between efficient radial transport and complete suppression. More recent observations of disks with large cavities reveal substantial diversity: some systems are strongly depleted in molecular emission, while others retain molecular abundances comparable to full disks \citep[e.g.,][]{Arulanantham2025, Rossi2025}. This diversity indicates that the presence of rings, gaps, or cavities does not uniquely determine the volatile content of the inner disk. Moreover, inner-disk molecular emission depends not only on the supply of icy material but also on local gas density, dust opacity, irradiation, and photochemistry, complicating any direct mapping between observed line strengths and radial transport efficiency.

In this context, our results provide a physical framework for interpreting these observations. We find that rings and gaps formed by non-ideal MHD processes do not act as strong barriers to inward dust migration. Instead, they redistribute solids and modulate their radial transport without severing the connection between the outer and inner disk regions. As a result, disks exhibiting prominent substructures of MHD origin may still sustain significant inward fluxes of volatile-bearing grains, even when their continuum morphology resembles that of disks with strong pressure traps.

When the substructures are produced by a massive planet, the regulation of inward transport becomes more pronounced. Nevertheless, even in this case, the gap does not behave as a perfectly sealed barrier. Gas and small grains can cross the gap through vertically extended flow channels, while larger grains may be partially filtered but not completely halted. 
Furthermore, the permeability of the gap evolves with time, with the resulting transport intrinsically azimuthally asymmetric and grain-size dependent.
%

These results suggest that interpreting JWST observations in terms of ``open'' versus ``closed'' transport barriers may be overly simplistic. Instead, the data are more naturally understood in terms of partial filtration of volatile-bearing solids. In particular, disks with reduced inner-disk water emission may not necessarily be completely cut off from outer-disk reservoirs; rather, they may experience diminished or selectively filtered transport, depending on the nature of their substructures. Conversely, the presence of substantial inner-disk volatile emission does not rule out the existence of gaps or even giant planets if those structures remain partially permeable.

Overall, our results indicate that outer-disk rings and gaps, whether of MHD or planetary origin, should be interpreted as regulators rather than absolute barriers of radial transport. This perspective provides a natural explanation for the diversity of molecular emission observed by JWST and highlights the need for caution when inferring the inner-outer disk connectivity or volatile delivery efficiency from dust continuum morphology alone.

\subsection{NC/CC meteoritic dichotomy}
\label{subsec:nccc}

Our results on gas and dust transport also have implications for interpreting the non-carbonaceous (NC)-carbonaceous (CC) meteoritic dichotomy, which is widely believed to reflect two spatially and chemically distinct reservoirs in the early Solar System \citep{Kruijer2017, Kruijer2020, Warren2011}. In the standard picture, these reservoirs were separated by the early formation of Jupiter, whose gap in the protoplanetary disk inhibited the radial mixing between the inner (NC) and outer (CC) regions \citep{Kruijer2017, Brasser2020}. This scenario implicitly assumes that a sufficiently massive planet can establish a long-lived and largely impermeable barrier. 

Our results suggest a more nuanced picture. Even in the presence of a Jupiter-mass planet capable of carving a deep gap, radial transport across the gap is not fully suppressed. Gas continues to cross the gap through vertically extended, azimuthally nonuniform flows, and small dust grains well coupled to the gas can follow these pathways. Larger grains are more strongly affected by pressure gradients at the gaps and tend to accumulate outside the planet, but filtration remains incomplete. The gap, therefore, acts as a size-dependent filter, reducing and delaying inward transport rather than completely stopping it.
Crucially, our simulations also show that a substantial population of solids, including relatively large grains, can already reside interior to the planet's orbit before the super-Keplerian gas motion near the outer edge of the planet-induced gap becomes strong enough to halt inward grain migration\footnote{We note that incomplete filtration across a Jovian gap has been discussed previously in viscous/turbulent disk models. In particular, \cite{Homma24} showed that selective leakage of dust across Jupiter’s gap, together with dust growth, fragmentation, and turbulent diffusion, can still be consistent with the NC/CC isotopic dichotomy while producing temporal isotopic variations. Their isotopic evolution calculations explicitly demonstrated that substantial, size-dependent dust leakage across Jupiter's gap remains consistent with the observed NC/CC isotopic dichotomy and the temporal evolution of meteoritic isotopic compositions. The new aspect emphasized here is not incomplete filtering by itself, but rather how such filtering operates in a three-dimensional, non-ideal MHD wind-launching disk, where radial transport is shaped by magnetic stresses, meridional circulation, and azimuthally intermittent accretion channels, without a prescribed viscosity.}. 
 We caution that the early pre-loading of large grains in the inner disk in our simulations is influenced by the idealized insertion of a fully formed giant planet into an initially undepleted disk and should not be interpreted literally as a realistic evolutionary timescale. Nevertheless, the results demonstrate the plausibility of substantial radial grain transport before a deep gap and efficient filtering barrier are fully established, particularly during earlier stages of planet growth when the planet mass is smaller.

In this framework, the NC/CC dichotomy does not require a perfectly impermeable Jovian barrier. Instead, it can arise from the combined effects of early mixing and subsequent size-dependent filtration. This scenario naturally allows the inner and outer reservoirs to become chemically distinct while still allowing limited communication between them. This may help explain why the NC and CC reservoirs are clearly separated in isotopic space, yet not completely disconnected.

Overall, our simulations support a reinterpretation of the Jovian barrier as a time-dependent and size-dependent filter operating on a disk that has already experienced significant early mixing. The meteoritic record may therefore reflect not a static and impermeable divider, but rather a transition from early mixing to later filtration.

\subsection{Pebble isolation mass in magnetic wind-launching disks}

Although our simulations focus on the global problem of radial transport of solids across a planet-induced gap, they inform the issue of local planetary core growth through pebble accretion within the Hill sphere via the concept of pebble isolation mass. In classical hydrodynamic models, it is defined as the mass of the planet at which a sufficiently strong pressure maximum forms outside the planetary orbit, stopping the inward drift of pebbles and thereby cutting off the supply of solids to the planet \citep[e.g.,][]{Lambrechts2014, Bitsch18}. For a disk with aspect ratio $h/r \sim 0.05$, this threshold is typically of the order $\sim 20$--$30\,M_\oplus$, depending on the level of turbulent diffusion and particle size. In this picture, pebble isolation is often treated as a relatively sharp transition in which radial pebble flux across the planetary orbit is effectively shut off.

Our results suggest that the above picture needs to be modified for non-ideal MHD wind-launching disks. In particular, the $0.1\,M_{\rm J}$ model (corresponding to $\sim 30\,M_\oplus$) lies near or above the classical pebble isolation mass, but it does not produce a persistent barrier to the inward transport of solids. Instead, the disk remains in a mixed regime in which wind-driven accretion flows coexist with planet-induced perturbations, and large grains continue to migrate inward across the outer edge of the relatively shallow gap. Even in the $1\,M_{\rm J}$ case, where a deep gap is formed, radial transport is not completely halted but proceeds intermittently through localized, azimuthally confined inflow channels.

We emphasize that pebble accretion onto the planetary core involves an additional step beyond the crossing of the outer gap edge: their subsequent capture by the planet depends on local gas dynamics and particle coupling within the Hill sphere, which are not resolved here. Within this limitation, our results show that the radial supply of solids is not abruptly terminated even for planets near or above the classical pebble isolation mass. Instead, planets impose a time-dependent and incomplete reduction in inward solid flux, with leakage persisting through localized, non-axisymmetric flow channels. Pebble isolation should therefore be understood not as a sharp cutoff, but as a flow-dependent filtering process whose efficiency is set by the global disk dynamics, including the nature of angular momentum transport and the origin of disk substructures.

\section{Conclusions} \label{sec:conclusion}

We have carried out three-dimensional non-ideal MHD simulations of wind-launching disks with embedded planets, including both gas-only and gas+dust models, to determine how magnetic stresses, planet-driven perturbations, and dust--gas coupling jointly regulate radial transport. Our main conclusions are as follows.

First, the dynamical impact of an embedded planet is strongly dependent on its mass and evolves over time. Low-mass planets leave the disk magnetic wind-dominated, with ring and gap substructures and fast midplane accretion streams set by magnetic stresses. As the mass of the planet increases, the system transitions through a mixed regime to a Jupiter-mass regime, where the planet opens a deep gap and drives strong spiral shocks. Even in this regime, the planet does not replace the wind-driven accretion system; it locally reshapes it, distorting the midplane current sheet, modifying meridional circulation, and redistributing magnetic flux, while vertically structured accretion flows persist. Neither MHD-generated rings and gaps nor planet-opened gaps act as absolute barriers to radial gas transport. Wind-driven substructures modulate, but do not halt inward flow. A Jovian planet can temporarily suppress or reverse transport near the outer gap edge at early times, but this weakens as the gap depletes. At later times, inward transport proceeds through localized, azimuthally intermittent channels, so the gap is intrinsically time dependent and partially permeable.

Second, dust transport is strongly size-dependent and closely coupled to the time-dependent, non-axisymmetric gas flow. Without a massive planet, small grains remain tightly coupled to the gas and trace its substructure, while larger grains settle toward the midplane and drift inward efficiently along fast accretion streams, crossing multiple MHD-generated rings and gaps before accumulating in the inner disk. Thus, wind-driven substructures redistribute solids but do not block their transport. In the presence of a Jupiter-mass planet, the behavior diverges by size: small grains continue to cross the gap with the gas, whereas larger grains are efficiently trapped at the pressure maximum outside the planet, forming prominent dust rings. Nevertheless, transport remains non-axisymmetric and time dependent, with both gas and dust crossing the gap through localized inflow channels. The planet, therefore, acts as an efficient but incomplete filter rather than a perfect barrier. In addition, the largest grains are significantly depleted in the circumplanetary disk, consistent with inefficient delivery through vertically structured accretion flows in the circumstellar disk.

Third, our results have direct implications for volatile delivery and the chemical evolution of planet-forming disks. Because radial transport is only partially suppressed, disk substructures regulate rather than halt the inward flux of solids and volatiles. In the context of JWST observations of inner-disk molecular emission, outer-disk icy material can be delivered inward both prior to gap opening and through continued, spatially intermittent leakage at later times, helping to explain the diversity of inferred inner-disk compositions. For Solar System formation, a giant planet can maintain a strong but imperfect separation between inner and outer reservoirs: early transport can preload the inner disk, while subsequent partial filtration and leakage are consistent with the persistence of the NC/CC dichotomy without requiring a completely impermeable barrier. Notably, even planets near or above the classical pebble isolation mass in hydrodynamic disks can still permit inward pebble transport in magnetically driven disks, indicating that the classical concept of pebble isolation requires modification in such environments, where radial transport remains only partially suppressed and the effective isolation threshold is shifted to higher planet masses.

\section*{Acknowledgements}

We thank the referee for a constructive report that improved the presentation of the paper, Pinghui Huang for help with the dust module in ATHENA++ used in this work, and Leo Krapp for useful discussions about embedded planets. CYH acknowledges support from the NRAO ALMA Student Observing Support (SOS) and computing resources from UVA research computing (RIVANNA), NASA High-Performance Computing, and NSF ACCESS (AST200032, PHY240192). ZYL is supported in part by NASA 80NSSC20K0533, NSF AST-2307199, JWST-GO-02104.002-A, JWST-GO-08872.003-A, and the Virginia Institute of Theoretical Astronomy (VITA). MKL is supported by the National Science and Technology Council (grants 113-2112-M-001-036-, 114-2112-M-001-018-, 113-2124-M-002-003-, and 114-2124-M-002-003-), an Academia Sinica Career Development Award (AS-CDA-110-M06), and an Academia Sinica Grand Challenge Seed Program (AS-GCS-115-M02). MKL thanks the Tsung-Dao Lee Institute for its hospitality, where part of this work was completed.

\section*{Data Availability}

The data underlying this article will be shared on reasonable request to the corresponding author.


\bibliographystyle{mnras}
\bibliography{ref} 




\appendix


\bsp	
\label{lastpage}
\end{document}